\documentstyle[12pt,fleqn,epsfig]{article}
\font\fontb=cmr12 scaled\magstep2 % 1.2*1.2
\font\fontc=cmr12 scaled\magstep1 % 1.2
\def\beq {\begin{eqnarray}}
\def\eeq {\end{eqnarray}}
\def\beqn {\begin{eqnarray*}}
\def\eeqn {\end{eqnarray*}}
\def\neqn {\nonumber}

\def\ni {\noindent}
\def\new {\newpage}
\def\PL #1 #2 #3 {Phys. Lett.~{\bf#1} (#2) #3}
\def\NP #1 #2 #3 {Nucl. Phys.~{\bf#1} (#2) #3}
\def\ZP #1 #2 #3 {Z.~Phys.~{\bf#1} (#2) #3}
\def\PR #1 #2 #3 {Phys. Rev.~{\bf#1} (#2) #3}
\def\PP #1 #2 #3 {Phys. Rep.~{\bf#1} (#2) #3}
\def\PRL #1 #2 #3 {Phys. Rev.~Lett.~{\bf#1} (#2) #3}
\def\PTP #1 #2 #3 {Prog. Theor.~Phys.~{\bf#1} (#2) #3}
\def\MPL #1 #2 #3 {Mod. Phys.~Lett.~{\bf#1} (#2) #3}
\def\IJM #1 #2 #3 {Int. J.~Mod.~Phys.~{\bf#1} (#2) #3}
\def\del{\partial}
\def\ra {\rightarrow}
\def\etal {{\it et al}.}
\def\eg {{\it e.g}.}
\def\ie {{\it i.e}.}

\def\bra {\langle}
\def\ket {\rangle}
\def\GeV{\mbox{GeV}}
\def\MeV{\mbox{MeV}}

\begin{document}
%%%%%%%%%%%%% PAGE %%%%%%%%%%%%%%
%\begin{titlepage}
\vspace* {-1cm}
%
%\begin{flushright}{9-5}
%\end{flushright}
%}
\vspace {0.5cm}
\baselineskip=0.8cm
\begin{center}{\fontb Chiral constituent quarks and  their role in 
quark distribution functions of nucleon and pion}
\footnote{This work is supported in part by BMBF} 
\\

\vspace {1.3cm}
{\fontc K. Suzuki\footnote{Alexander von Humboldt fellow, 
e-mail address : ksuzuki@physik.tu-muenchen.de} and W. Weise\\
{\em Physik-Department, Technische Universit\"{a}t M\"{u}nchen,}\\ 
{\em Theoretische Physik,}\\
{\em D-85747 Garching, Germany}}

\vspace {1.8cm}
{\bf Abstract}
\end{center}

\baselineskip=0.78cm
\noindent
We investigate the structure of constituent quarks and study 
implications for 
quark distribution functions of hadrons.  Constituent quarks are constructed 
by dressing bare quarks with Goldstone bosons using the chiral quark
model.  We calculate resulting corrections to the twist-2 structure functions 
$f_1(x)$, $g_1(x)$ and $h_1(x)$.   
The Goldstone boson fluctuations produce a flavor asymmetry of the quark
distribution in the nucleon in agreement with experimental data.  
They also generate significant depolarization effects which reduce the
fraction of the nucleon spin carried by quarks.   
Corrections to the transversity spin structure function $h_1(x)$ 
differ from those to $g_1(x)$, and  
in particular  we find a large reduction ($40\%$) of the $d$-quark tensor 
charge, which is consistent with recent lattice calculations.  
We also study the pion structure function and 
find the momentum fraction carried by the sea quarks in the pion to be   
considerably larger than that in the nucleon.

\vspace{0.6cm}
\baselineskip = 0.7cm

\ni

\ni

%\end{titlepage}

%-------------%-------------%-------------%-------------%-------------
\newpage
\baselineskip = 0.765cm

\ni
{\bf 1 Introduction}

\ni
High energy experiments such as lepton-hadron deep inelastic scattering
and the Drell-Yan process provide detailed knowledge of the quark and gluon 
distributions of hadrons.  
One obtains information about the momentum distribution functions 
of valence quarks, sea quarks and gluons, and their scaling
violations are quite consistent with predictions from perturbative 
QCD \cite{Muta}.  
Even under these circumstances, we still have a most important problem, 
namely how to understand the structure functions themselves.    
Perturbative QCD  describes their $Q^2$ evolution, but cannot
predict the distribution functions because of 
their non-perturbative origin.   
Although lattice QCD studies provide some moments of the structure 
functions \cite{Lattice}, a satisfactory understanding is still far from 
being reached.

On the other hand, much of low-energy hadron phenomenology is quite
successfully described in terms of the constituent
quark picture.  At low energies, the sea quark and gluon degrees of freedom
are assumed to be absorbed into  constituent quarks as quasi-particles, 
and hadrons are constructed out of a few such quasi-particles.  
At the high energy scale, sea quarks and gluons reappear and hadrons 
reveal themselves as complex many-body systems with a large number of 
current quarks and gluons.
We do not yet have a clear understanding of the connection between 
constituent quarks at low energy and the parton picture at high energy.

Some attempts have been made to model this connection or at least some aspects
of it \cite{Jaffe,Bag,Kulagin,models}.   
In such studies, the twist-2 part of a given structure function is
evaluated within a relativistic quark model, and the resulting structure
function is then 
evolved from the low-energy scale, where the constituent quark picture is
supposed to work, to the high momentum scale with the help of  
perturbative QCD.  
Such studies may help to approach the deep
inelastic scattering data from the point of view of low-energy 
non-perturbative dynamics.   
For example, the observed large deviations of the ratios 
$F^n_2(x) / F^p_2(x)$ and $g_1^p(x) / F^p_1(x)$ from the simple 
parton picture can be  understood by taking into
account the spin-flavor structure of the nucleon \cite{Close_Thomas,scalar} 
in terms of the structure of constituent quarks.    
In addition, the difference 
between the $u$-quark distribution in the pion and the one in the kaon is well
reproduced within a chiral model \cite{Shigetani}  
where the constituent quarks play a crucial role.

Despite its successes, the perturbative QCD evolution becomes problematic 
when it is started from scales far below 1 GeV.  
In most cases, calculations require that the evolution is performed 
upward from scales as low as 
$\mu ^2 \sim 0.1 - 0.3 \GeV^2$ to reproduce the high $Q^2$ experimental data.  
In this region the use of  perturbative QCD may be questionable,   
even though the difference between results obtained in leading order
and next-to-leading order from a scale $\mu^2 \sim 0.3 \GeV^2$ is 
only about $10\%$.

Recent studies of Kulagin {\etal}\cite{Kulagin} have shown that 
such difficulties can be overcome 
by taking into account the structure of the constituent quarks.  
It was found that inclusion of the pion dressing and 
higher mass spectator processes substantially
changes the shapes of the quark distributions at the given model scale.  
This modifies the normalization of the quark distributions 
and leads to the correct
small-$x$ behavior by introducing Regge exchange.   
The identification of the results in ref.~\cite{Kulagin} with the twist-2 
quark distribution can now be done at a scale of about 
$1\GeV^2$, where the use of perturbative QCD for further evolution to high
$Q^2$ is reasonably justified.

In the present paper we follow a similar direction, although with 
different emphasis, and extend it to the 
spin dependent structure functions and to hadrons other than the nucleon.  
At the scale below 1 GeV, the relevant degrees of freedom are assumed to be 
constituent quarks (CQ) and  Goldstone (GS) bosons\cite{CQM}.  
Here, we concentrate on the Goldstone boson
dressing in the $SU(3)_R \otimes SU(3)_L$ 
chiral quark model with explicit $SU(3)$ flavor symmetry breaking.     
We study the twist-2 structure functions of the nucleon \cite{JaffeJi},
namely the unpolarized $f_1(x)$, the helicity difference $g_1(x)$, and 
the transversity difference $h_1(x)$,  as well as the pion structure function 
with inclusion of the CQ structure.  
While Regge high-energy behaviour was found to be important in previous studies
of the unpolarized structure function, primarily at small $x$, our main focus 
here will be on the ``soft" dynamics mediated by the pseudoscalar meson 
octet.
The GS boson cloud influences not only the normalization at the quasi-particle 
pole of the constituent
quarks but also changes their spin structure by emitting the GS bosons into 
$P$-wave states relative to the CQ, thus producing depolarization effects.

Recently, such a pion dressing has been studied by several 
authors \cite{Eichten} in order to account for 
the violation of the 
Gottfried sum rule and the nucleon spin structure.   
We first reexamine this previous work systematically.  
%Calculated result for the Gottfried sum with the standard parameter of the
%model is consistent with the experimental data, but the nucleon spin is found
%to be large compared with the observed value $\Sigma \sim 0.3$.  
%The GS boson dressing can not reproduce the nucleon axia-vector coupling $G_A$
%and the nuclen spin $\Sigma$ simulattaneously.  
Then we study the chiral-odd transversity spin structure function 
$h_1(x)$.  
Corrections to $h_1(x)$ turn out to differ from those for $g_1(x)$.  
This difference causes a strong reduction of the $d$-quark tensor charge, 
while 
the axial charge remains almost unchanged.  
This tendency seems to be consistent with a recent lattice 
study \cite{h1lattice} which suggests $|\delta d| < |\Delta d |$ as 
opposed to $\delta u > \Delta u$ for the $u$-quarks.

We also show how the pion structure function modifies with 
inclusion of CQ structure, in comparison with the nucleon.  
We find that the resulting sea quark distribution in the pion is 
substantially enhanced.  The contribution to the second moment from the 
sea quarks in the pion is almost twice as large as that in the nucleon.  
This issue will possibly be studied in future experiments.

This paper is organized as follows.  In Section 2, we construct the constituent
quark Fock states and evaluate their contributions to  deep inelastic 
scattering.   
We examine the Gottfried sum rule and the nucleon spin structure in Section 3, 
as it was 
first done by Eichten {\etal}\cite{Eichten}.  
In Section 4 we study the chiral-odd transversity spin structure function
$h_1(x)$ and show that the GS boson dressing of quarks changes the simple 
quark model result for $h_1^d(x)$ substantially.  
Section 5 is focused on  the discussion of the pion structure function.  
We emphasize that the inclusion of the GS boson dressing naturally
leads to an enhancement of the sea quarks in the pion.  
We draw conclusions in the final section 
with a brief estimate of  
contributions from multi-pion Fock states.

\vspace{2cm}

%-----------%-----------%-----------%-----------%-----------%-----------%
%\new
\ni
{\bf 2 Structure of  constituent quarks}

\ni
We start by constructing the constituent quark Fock state   
using the chiral quark model of Manohar and Georgi \cite{CQM}.  In this 
model, 
constituent quarks couple to the Goldstone bosons of  
spontaneously broken chiral symmetry.  The GS bosons, in 
particular the pion, play a 
crucial role as approximate zero modes of the QCD vacuum.  
They govern the low-energy dynamics at characteristic scales 
$\mu {^< _\sim} 4 \pi f_\pi \sim   1 \GeV$ 
where $f_\pi \simeq 93 \MeV$ is the 
pion decay constant.

Let $\psi = (u,d,s)$ be the quark field with $N_f = 3 $ flavours.  
The effective interaction Lagrangian in leading order is given by
\beq
{\cal L}_{\mbox{int}} = - \frac {g_A} {f} \bar \psi  \gamma^{\mu} 
\gamma^5  \left( \partial_{\mu}  \Pi  \right) \psi
\label{lagrangian}
\eeq
with the GS boson matrix field,
\beq
\Pi ={1 \over {\sqrt 2}}\left( {\matrix{ \frac{\pi^0}{\sqrt{2}} + 
\frac{\eta}{\sqrt{6}}  &\pi^+ &K^+ \cr
\pi^- & - \frac{\pi^0}{\sqrt{2}} + \frac{\eta}{\sqrt{6}} &{K^0}\cr
K^- & \bar K^0 & - \frac{2\eta}{\sqrt{6}} \cr
}} \right) \;\; .
\eeq
Here, we set the pseudoscalar decay constant $f$ equal to the pion decay 
constant $f_\pi$  and start with 
the quark axial-vector coupling $g_A = 1$.  
In their original work, Manohar and Georgi took $g_A = 0.75$ to reproduce 
the nucleon axial-vector coupling constant
within the non-relativistic approximation\cite{CQM}.   
We first adopt $g_A$ = 1 following the large $N_c$ argument from 
ref.~\cite{Weinberg} and discuss possible renormalization effects later.

The $u$-, $d$- and $s$-quarks that enter in eq.~(\ref{lagrangian}) are assumed 
to have already developed large dynamical 
masses as  a dynamical 
consequence of spontaneous chiral symmetry breaking.  We denote 
those ``bare'' but massive states by $\left| u_0 \right\rangle$ and 
$\left| d_0 \right\rangle$ etc.  
Once they are dressed by GS bosons, 
we write the constituent $u$- and $d$-quark Fock-states as 
{
\setcounter{enumi}{\value{equation}}
\addtocounter{enumi}{1}
\setcounter{equation}{0}
\renewcommand{\theequation}{\theenumi-\alph{equation}} 
\beq
\left| U \right\rangle  &=&\sqrt{Z}\left| u_0 \right\rangle + 
a_\pi \left| {d \pi ^+} 
\right\rangle +{{a_\pi } \over 2}\left| {u \pi ^0} \right\rangle +a_K\left| 
{s K^+} \right\rangle +{{a_\eta } \over 6}\left| {u \eta }
 \right\rangle   \,\, ,
\label{u-fock}
\eeq
\beq
\left| D \right\rangle &=& \sqrt{Z}\left| d_0  \right\rangle +a_\pi 
\left| {u \pi ^-} 
\right\rangle +{{a_\pi } \over 2}\left| {d \pi ^0} 
\right\rangle +a_K\left| {s  K^0} \right\rangle +{{a_\eta } 
\over 6}\left| 
{d \eta } \right\rangle \,\, , 
\label{d-fock}
\eeq
\setcounter{equation}{\value{enumi}}
}

\vspace{-0.5cm}
\ni
where $Z$ is the renormalization constant for a ``bare''
constituent quark and $| a_i | ^2$ are probabilities to find GS bosons in the 
dressed constituent quark states.   
The wave function renormalization $Z$ only operates on 
the bare CQ state, because 
we use the renormalized quark-meson coupling constants to calculate 
$a_i$.  
In this paper we restrict our study to the admixture of one 
GS boson and truncate the Fock space expansion as displayed in 
eqs.~(\ref{u-fock}), (\ref{d-fock}).  
We shall give an  estimate of two-pion Fock state contributions later.

Within this approximation, the diagrams Figs.~1(a) and 1(b) contribute to the 
structure functions.  
We use the Infinite Momentum Frame (IMF) to calculate these
contributions\cite{Drell,Review_IMF}.  
By virtue of the IMF,  the factorization of the subprocess is
automatic and we may neglect possible off-shell corrections, since all the 
particles are on-mass-shell in this frame.  
Hence we can use one-dimensional convolution formalism throughout the 
following calculations.   Also, working in the 
IMF removes the so-called Z-graph contributions.

%%%%%%%%%%%%%%%%%%%%%%%%%%%%%%%%%%%%%%%%%%%%%%%%%%%%%%%%%%%%%%%%%%%%%%%%
\begin{figure} 
\begin{center} 
\psfig{file=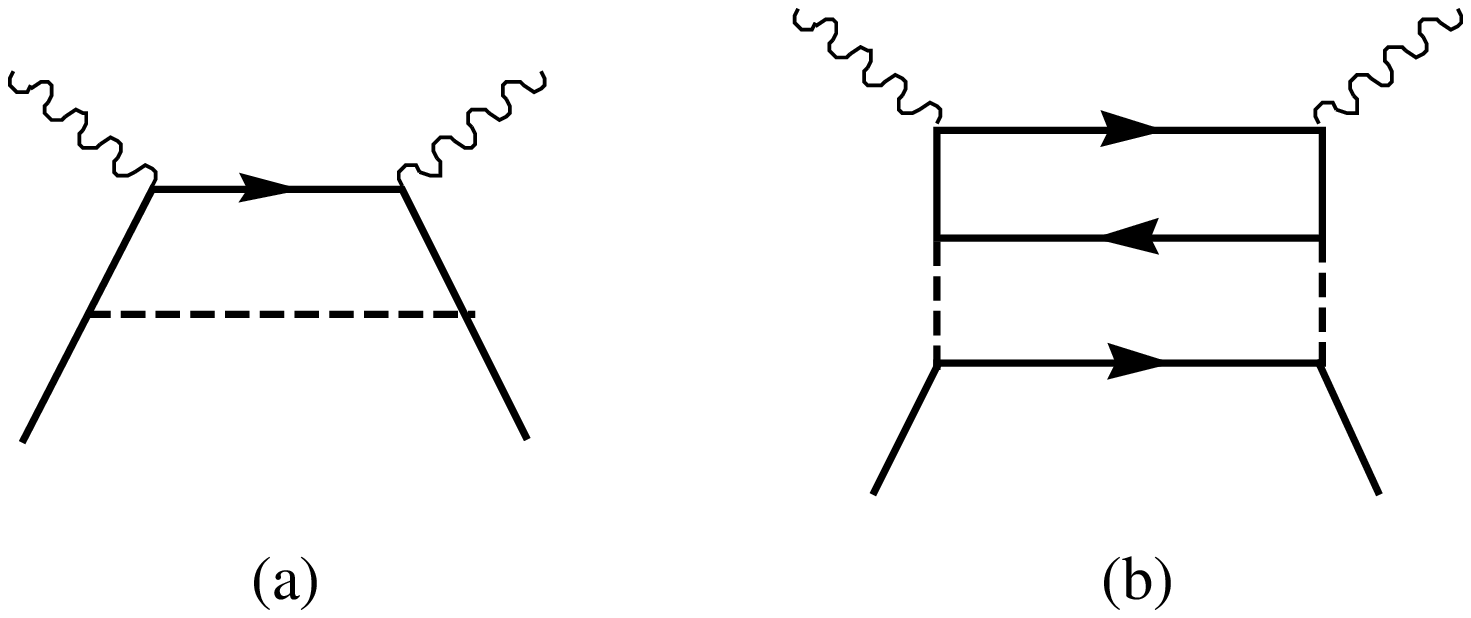,height=1.6in} 
\caption{}   
\end{center} 
\vspace{-0.2cm} 
Diagrams contributing to the constituent quark structure.  
Fig.1.(a) shows the GS boson spectator process, whereas Fig.1(b) probes 
the structure of the GS boson itself, with the constituent quark being
spectator.  
The virtual photon is depicted by the wavy line.  
Thick sold curves represent the quarks, and the dashed curve the
GS boson.  
\end{figure} 
%%%%%%%%%%%%%%%%%%%%%%%%%%%%%%%%%%%%%%%%%%%%%%%%%%%%%%%%%%%%%%%%%%%%%%%%

In previous studies we have already obtained the pion dressing corrections
to the constituent quark distribution function at a low-momentum 
scale \cite{Kulagin}, as follows.    
The spin-independent term corresponding to diagram 1(a) is given by
\beq
q_j (x) =
\int_x^1 {{dy} \over y} \,P_{j \, \alpha / i} (y) \, q_i \left( {{x \over y}} 
\right) \;\; . 
\eeq
Here, 
$P(y)_{j \, \alpha / i}$ is the 
splitting function which gives the probability to find a constituent quark
$j$ carrying the light-cone momentum fraction $y$ together with a 
spectator GS boson $(\alpha = \pi,K,\eta)$, both of which coming from 
a parent constituent quark $i$: 
\beq
P(y)_{j \, \alpha / i}={1 \over {8\pi ^2}}\left( {{{g_A \, \bar m} 
\over {f }}} \right)^2\int_{}^{} {dk_T^2}
{{(m_j - m_i y)^2 + k_T^2} \over {y^2 (1-y) 
\left[ {m_i^2-M^2_{j \alpha}} \right]^2}} \;\; , 
\label{spil-f}
\eeq
where $m_i, m_j, m_{\alpha}$ are 
the masses of the $i,j$- constituent quarks and 
the pseudoscalar meson $\alpha$, respectively.  
$M^2_{j \, \alpha } = \frac{m_j^2+k_T^2}{y} 
+ \frac{m_{\alpha}^2 + k^2_T}{1-y}$ is the invariant mass squared of the final
state, and $\bar m$ is the average of the constituent quark masses, 
$\bar m = (m_i + m_j) / 2$.   The integral (\ref{spil-f}) requires a 
cutoff which will be specified later.

Diagram 1(b) probes the internal structure of the GS bosons.  
This process gives the following contribution: 
\beq
q_k (x)  =
\int {{{dy_1} \over y_1}}  \, {{{dy_2} \over y_2}}
\, V_{k/ \alpha} \left( \frac{x}{y_1} \right) P_{\alpha  \, j / i} 
\left( \frac{y_1}{y_2} \right) \,q_i (y_2)  \, , 
\eeq
where $P_{\alpha \, j / i} (x) = P_{ j \, \alpha / i} (1-x)$.   This 
symmetry relation holds in the IMF if we use a
momentum cutoff procedure as in ref.~\cite{Review_IMF}.  
Here, $V_{k / \alpha }(x)$ is the quark distribution function with flavor $k$ 
in the GS boson $\alpha$, with the normalization $\int^1_0 {dx} V_{k / 
\alpha }(x)  =1$.  
When we calculate the quark distribution explicitly in
Section 4 and 5, we will use the phenomenological parametrization 
of the pion structure function at the scale $1 \GeV^2$\cite{Gluck_pi}.  
As for the kaon and eta, we do not have 
experimental data and simply use the model calculations of 
ref.~\cite{Shigetani,pi_NJL}.  Ambiguities of the final results arising from 
the choice of the meson structure functions are rather small, at the 
level of a few $\%$.

%When we calculate the quark distribution explicitly in
%section 4 and 5, we will use results of the model calculations of 
%ref.~\cite{Shigetani,pi_NJL}, which reproduce the experimental data and 
%are consistent with the phenomenological parametrization of 
%ref.~\cite{Gluck_pi}.   
%We include all the octet GS bosons in the following 
%studies, although the pion contributions are expected to dominate.  

We define the moments of the splitting functions $P_{ j \alpha, / i}$, 
\beq
\left\langle  x ^{n-1} P_{j \, \alpha / i } \right\rangle \equiv 
\int_0^1 {dx}\;x^{n-1}\;P_{j \, \alpha / i } (x) \; \; .  
\eeq
As for the first moments, $\bra P_{j \, \alpha / i } \ket = 
\bra P_{\alpha \, j / i } \ket \equiv \bra P_\alpha \ket $.  
In terms of those the renormalization constant $Z$ is then given by 
\beq
Z=1 -  {3 \over 2}\left\langle {P_\pi } \right\rangle  -\left\langle 
{P_K} \right\rangle -{1 \over 6}\left\langle {P_\eta } 
\right\rangle  \;\; .
\label{Z}
\eeq
We find $Z \sim 0.7$ using the standard parameter set $g_A=1$, $\bar m = 350$ 
MeV and a cutoff $\Lambda = 4 \pi f$ as specified at the end of this Section.  
Numerically, about 75$\%$ of the deviation from $Z=1$ 
comes from the pion dressing and 20 $\%$ from the kaons.

%-------%-------%-------%-------%-------%-------%-------

As examples, we explicitly write down the quark distribution functions in the 
nucleon using the splitting functions (\ref{spil-f}).  
We assume that the bare quark distribution functions are given in terms of the 
valence quark distributions $u_0(x)$ and $d_0(x)$. No antiquarks are 
present before the GS boson coupling is turned on.  We find 
\beq
u(x)&=&Z u_0(x)+P_{u \pi / d} \otimes d_0 + 
V_{u / \pi }\otimes P_{\pi \, d / u} \otimes u_0+
{1 \over 2}P_{u \, \pi/ u} \otimes u_0  \neqn \\
 & &+ {1 \over 4}V_{u / \pi }\otimes P_\pi \otimes (u_0+d_0)+ 
V_{u / K}\otimes P_{K \, u / u }\otimes u_0 \neqn \\
& & + \frac{1}{6} P_{u \, \eta/ u} \otimes u_0 + {1 \over 36}
V_{u / \eta }\otimes P_\eta \otimes (u_0+d_0) \;\; ,
\label{u_N}
\eeq
\beq
d(x)&=&Z d_0(x)+P_\pi \otimes u_0+V_{d / \pi }\otimes P_\pi \otimes d_0+
{1 \over 2}P_\pi \otimes d_0  \neqn \\
 & &+ {1 \over 4}V_{d / \pi }\otimes P_\pi \otimes (u_0+d_0)+ 
V_{d / K}\otimes P_K\otimes d_0 \neqn \\
& & + \frac{1}{6} P_\eta \otimes d_0 + {1 \over 36} 
V_{d / \eta }\otimes P_\eta \otimes (u_0+d_0) \;\; , 
\label{d_N}
\eeq
where the bare distribution functions in the proton have the normalizations
\beqn
\int^1_0 {dx} \, u_0(x) = 2 \, ,\\
\int^1_0 {dx} \, d_0(x) = 1 \, .
\eeqn
Here, we use the following short hand notation for the convolution integral: 
\beq
P\otimes q\equiv \int_x^1 {{{dy} \over y}} \, P(y) \,\, 
q\left( {{x \over y}} \right) \; .
\eeq
For the antiquarks and the strange quarks we find: 
\beq
\bar u(x)&=& V_{u / \pi} \otimes P_\pi \otimes d_0  \neqn \\
&& + \frac{1}{4} V_{u / \pi} \otimes P_\pi \otimes (u_0+d_0) +
\frac{1}{36} V_{u / \eta} \otimes P_\eta \otimes (u_0+d_0) \; ,
\label{baru_N}
\eeq
\beq
\bar d(x)&=& V_{d / \pi} \otimes P_\pi \otimes u_0 \neqn \\
&& + \frac{1}{4} V_{d / \pi} \otimes P_\pi \otimes (u_0+d_0) +
\frac{1}{36} V_{d / \eta} \otimes P_\eta \otimes (u_0+d_0) \; ,
\label{bard_N}
\eeq
\beq
s (x)&=&  P_K \otimes (u_0 + d_0)+ \frac{4}{9} 
V_{s / \eta} \otimes P_\eta \otimes (u_0+d_0) \; ,\\
\bar s (x)&=&  V_{s / K} \otimes P_K \otimes (u_0 + d_0)+ \frac{4}{9} 
V_{s / \eta} \otimes P_\eta \otimes (u_0+d_0) \; .
\label{s_N}
\eeq
We then obtain the valence quark distributions as
$u_{val} = u(x) -\bar u(x),$ $ d_{val} (x) = d(x) - \bar d(x)$, 
which satisfy the correct normalization with the renormalization constant 
$Z$.  
For instance, the first moment of the valence $u$-quark is  
\beqn
\int ^1_0{dx} u_{val} (x) &=& \int^1_0{dx} 
\left[ Z u_0(x) + P_{u \pi /d} \otimes d_0 + 
V_{u / \pi }\otimes P_{\pi \, d / u} \otimes u_0  \right. \\ 
& & \left. + {1 \over 2} P_{u \, \pi / u} \otimes u_0  
-  V_{u / \pi }\otimes P_{\pi \, u / d} \otimes d_0 \right. \\ 
& &\left. + V_{u / K} \otimes P_{K \, u / u } \otimes u_0 + 
\frac{1}{6} P_{u \, \eta/ u} \otimes u_0 \right] \\
&=& 2 Z + \bra P_\pi \ket + 2 \bra P_\pi \ket + 2 \frac{1}{2} \bra 
P_\pi \ket - \bra P_\pi \ket + 2  \bra P_K  \ket 
+  2 \frac{1}{6} \bra P_\eta \ket  \\
&=& 2 \;\; ,
\eeqn
where we have used eq.~(\ref{Z}) in the last step.

%--------%--------%--------%--------%--------%--------%--------%--------

Next we evaluate the spin-dependent process and 
study the spin distribution of the nucleon.  
An analogous calculation as the one leading to eq.~(\ref{spil-f}) yields: 
\beq
\Delta P(y)_{j \, \alpha / i}={1 \over {8\pi ^2}}\left( {{{g_A \bar m} 
\over {f  }}} \right)^2\int {dk_T^2}
{{(m_j - m_i y)^2 - k^2_T } \over {y^2 (1-y) 
\left[ {m_i^2 - M^2_{j \, \alpha}} 
\right]^2}}  \; \; ,
\label{spil-g}
\eeq
where $\Delta P_{j \, \alpha / i}= P_{j\uparrow \, \alpha / i \uparrow } - 
P_{j \downarrow 
\, \alpha / i \uparrow }$ is the difference of probabilities 
 to find  helicity $+1/2$ quarks 
minus helicity $-1/2$ quarks starting from a parent quark with 
helicity $+1/2$.   Note the change of sign in front of $k_T^2$ when comparing 
$P$ and $\Delta P$.  
With the standard parameter set mentioned above, this spin dependent 
splitting function gives a negative contribution, that is, the helicity-flip 
process is dominant.  
In previous studies\cite{Eichten}, 
the relation $P= - \Delta P$ has simply been used, 
assuming that the GS boson emission contributes only to the helicity 
flip process.  
However, the ratio of helicity flip and non-flip contributions depends on the 
dynamics.  
The magnitude of the spin-dependent splitting function is rather sensitive to the 
choice of the cutoff, because the helicity-flip probability is directly
proportional to the transverse momentum integral.  
The GS boson dressing given by the formula (\ref{spil-g}) modifies the 
axial-vector coupling constant $g_A$ of the nucleon as well as 
the quark content $\Sigma$ of the nucleon spin, as we shall elaborate.

Now we specify the momentum cutoff function at the quark-GS boson vertex.
An  exponential cutoff is often used in IMF
calculations, since such a form factor has the correct $t$ and $u$ 
channel symmetry\cite{Review_IMF}:
\beq
g_A \ra g_A \;\, \mbox{exp} \left[ \frac{m_i^2 -M^2_{j \, \alpha }} 
{4 \Lambda^2} \right]  \; .
\label{cutoff-f}
\eeq
Here $\Lambda$ is the cutoff parameter.  This function satisfies the 
proper symmetry, $P_{j \alpha / i} (y) = P_{\alpha j / i} (1-y)$.  
The value of the cutoff is taken to be 
about $ \Lambda \sim 4 \pi f_\pi \sim 1 \GeV$, 
the characteristic scale of spontaneous chiral
symmetry breaking.   In actual calculations it will be determined to 
reproduce the experimental data of the Gottfried sum rule.

Using this cutoff function, we show in Fig.2 the constituent quark-Goldstone 
boson splitting functions, $P_{j \alpha /i} (y)$  and 
$\Delta P_{j \alpha /i} (y)$,   
for the case of a $u$-quark $(i=j=u)$ together with the pion 
$(\alpha = \pi)$  and $\Lambda = 1.4 \GeV$.    
For the constituent quark masses,
we use $m_u = m_d = 360 \MeV$ and $m_s = 570 \MeV$ as typical values 
guided by NJL model calculations\cite{NJL} and phenomenology.  
The resulting splitting function $P_{u \alpha /u}$ 
is peaked at $y < 0.4$, and    
$\Delta P_{j \alpha /i}$ is negative over most of the $y$ range.  
If we choose smaller values for the cutoff $\Lambda$ (say, 
$\Lambda = 0.5 \GeV$),  
$\Delta P_{j \alpha /i}$ becomes positive but stays very small.

%%%%%%%%%%%%%%%%%%%%%%%%%%%%%%%%%%%%%%%%%%%%%%%%%%%%%%%%%%%%%%%%%%%%%%%%
\begin{figure} 
\begin{center} 
\psfig{file=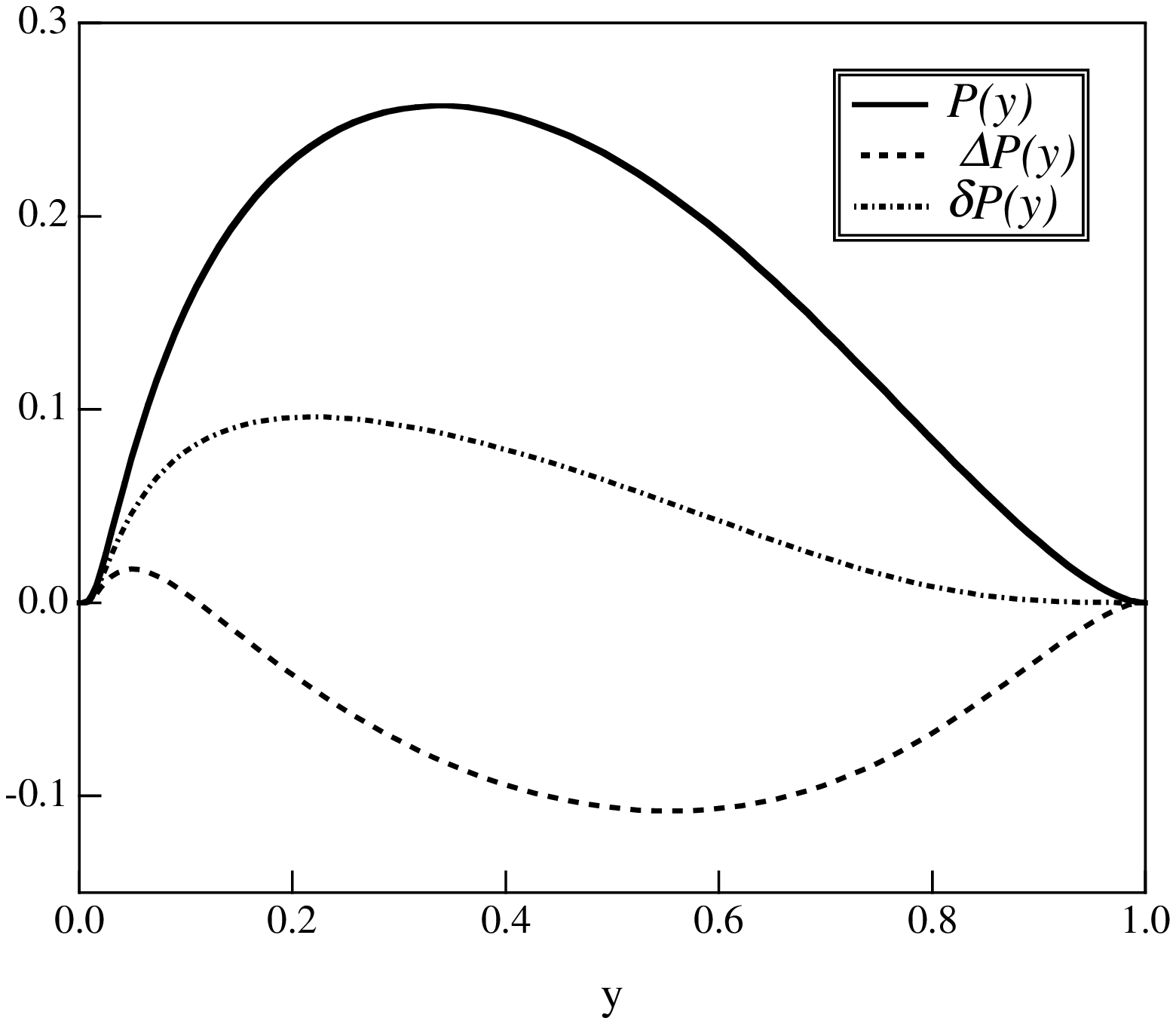,height=4in} 
\caption{}   
\end{center} 
\vspace{-0.2cm} 
Constituent quark-GS boson splitting functions for twist-2 structure
functions.  
$P_{j {\pi} /  i} (x)$,  $\Delta P_{j {\pi} /  i} (x)$ and 
$\delta P_{j {\pi} /  i} (x)$ 
are shown by the solid, dashed and dash-dotted curves, respectively.  
\end{figure} 
%%%%%%%%%%%%%%%%%%%%%%%%%%%%%%%%%%%%%%%%%%%%%%%%%%%%%%%%%%%%%%%%%%%%%%%%

%%%%%%%%%%%%%%%%%%%%%%%%%%%%%%%%%%%%%%%%%%%%%%%%%%%%%%%%%%%%%%%%%%%%%
\new
\ni
{\bf 3 Gottfried sum and nucleon spin in the chiral quark model\\
3.1 Update of the Gottfried sum rule}

\ni
We start this Section by re-examining the 
effects of the GS boson fluctuations on the Gottfried sum rule and 
the nucleon spin structure.  These were first discussed by Eichten 
{\etal}\cite{Eichten}.  
The Gottfried sum rule (GSR) 
is given in terms of the difference of the proton and neutron 
structure functions:  
\beq
GSR=\int_0^1 {\frac{dx}{x}\;}\left[ {F_2^p(x)-F_2^n(x)} \right]=
{1 \over 3}\int ^1_0 {dx} \left( {u(x)+\bar u(x)-d(x)-\bar d(x)} \right) \; .  
\label{gsr}
\eeq
A naive parton model with isospin symmetric sea, $\bar u = \bar d$, yields 
$GSR$ = 1/3, whereas the  
experimentally observed value is $0.235 \pm 0.026$\cite{NMC}, 
which indicates isospin symmetry breaking $\bar u(x) < \bar d(x) $ in the  
nucleon sea.  
If we take into account the shadowing correction to extract the neutron
structure function from the deuteron data, the GSR value is further reduced by 
about 20-30$\%$ \cite{Shadowing}.  With this correction, 
the empirical value of the GSR is reduced to about $0.21-0.22$.

We now employ the quark distribution functions of the chiral quark model
as given in Section 2.  Inserting the expressions 
(\ref{u_N},\ref{d_N},\ref{baru_N},\ref{bard_N}) into eq.~(\ref{gsr}), 
one gets 
\beq
GSR&=&{1 \over 3}\left( {Z-{1 \over 2}\left\langle {P_\pi } 
\right\rangle +\left\langle {P_K} \right\rangle+
{1 \over 6}\left\langle {P_\eta } \right\rangle } \right)\neqn \\
&=&{1 \over 3} \left( 1 - 2 \left\langle {P_\pi } 
\right\rangle  \right) \; .  
\label{gsr2}
\eeq
With the parameters discussed in Section 2, the renormalization constant is
found to be $Z \sim 0.7$.  
Here, we adopt $\Lambda = 1.4 \GeV$ for the cutoff function 
eq.~(\ref{cutoff-f}), and obtain $Z = 0.67$, 
$\left\langle {P_{\pi} } \right\rangle = 0.16$, 
$\left\langle {P_{K }} \right\rangle =   0.09$ and
$\left\langle {P_\eta } \right\rangle = 0.06$.  
We then find 0.22 for $GSR$, in good agreement with the 
empirical value.   
The dominant contribution to the reduction of GSR comes from the 
renormalization of the bare quark state represented by the $Z$ factor.  
The value $\Lambda = 1.4 \GeV$ should be understood as an upper 
bound of the cutoff in the chiral quark model, based on the assumption that 
the violation of the Gottfried sum is entirely given 
by the Goldstone boson dressing alone.   One should note, of course, that 
this $\Lambda $ depends on the input constituent quark mass.  
If we choose $\bar m \simeq 400 \MeV$, the resulting cutoff turns out to be 
about $\Lambda \simeq 1.1 \GeV$.

%If we take  $g_A= 0.75$ in the effective Lagrangian, then 
%a too large momentum cutoff $\ge 2 \GeV$ is needed to obtain the 
%reasonable GSR value $\sim 0.22$\cite{Eichten}.  
%This cutoff is much larger than the value $1 \GeV$ which is frequently 
%referred  as the chiral symmetry breaking scale.  
%With more realistic parameters, the cutoff $1.5 \GeV$ and $g_A=1$, 
%we can reproduce the data well.  

\vspace{2cm}

\new
\ni
{\bf 3.2 Angular momentum transfer to the meson cloud}

\ni
Next we discuss the spin structure of the nucleon.  
Analysis of all available experimental data\cite{Ellis} 
gives the following decomposition of the nucleon spin in terms of the quark
spin, with 
$\Delta q = 
q\uparrow -q\downarrow +\bar q \uparrow -\bar q\downarrow$:
\beq
&& \Delta u = 0.82 \pm 0.02, \hspace{1cm} \Delta d = -0.43 \pm 0.02, 
\hspace{1cm} \Delta s = -0.10 \pm 0.02,     \neqn \\
&& \Sigma  \equiv  \Delta u + \Delta d + \Delta s = 0.29 \pm 0.06
\label{spin_exp}
\eeq
at $Q^2 = 5 \GeV^2$,   
to be compared with expectations of the naive quark models ($\Sigma = 1$) or 
with inclusion of relativistic corrections $(\Sigma \sim 2/3)$.  
Using the spin-dependent splitting
functions (\ref{spil-g}), the spin fractions of the constituent quarks 
are modified from their bare quark values as follows: 
\beq
\Delta u = Z \Delta u_0 + \frac{1}{2} \left\langle 
 {\Delta P_\pi } \right\rangle \Delta u_0 + 
 \left\langle {\Delta P_\pi } \right\rangle  \Delta d_0 + 
\frac {1}{6} \left\langle {\Delta P_\eta }
\right\rangle  \Delta u_0 \; ,
\label{su}
\eeq
\beq
\Delta d = Z \Delta d_0 + \frac{1}{2} \left\langle 
{\Delta P_\pi } \right\rangle \Delta d_0 + 
 \left\langle {\Delta P_\pi } \right\rangle  \Delta u_0 + 
\frac {1}{6} \left\langle {\Delta P_\eta }
\right\rangle \Delta d_0 \; ,
\label{sd}
\eeq
\beq
\Delta s = \left\langle 
{\Delta P_K } \right\rangle  ( \Delta u_0 + \Delta d_0) \; ,
\label{ss}
\eeq
where we have assumed  $\Delta s_0  = 0$ at the moment.  
Recall that the first moments $\bra \Delta P \ket$ of the 
spin-dependent splitting functions 
$\left\langle {\Delta P_\pi } \right\rangle $ are negative, {\ie}~the 
spin-flip probability is larger than the spin non-flip one in the pion
emission process.  
We find 
$\left\langle {\Delta P_\pi } \right\rangle = -0.06$, 
$\left\langle {\Delta P_K } \right\rangle = -0.01$, and
$\left\langle {\Delta P_\eta } \right\rangle = -0.03$.  
The pion emission process converts part of the spin of the constituent quark 
into (P-wave) angular momentum of the meson cloud.

In order to obtain numerical results,  initial input spin fractions 
$\Delta u_0$ and $\Delta d_0$ are needed.   
As a first rough estimate, we start from the naive $SU(6)$ quark model values, 
$\Delta u_0 = {\frac{4}{3}}$, $\Delta d_0 = - \frac{1}{3}$, 
which yield the nucleon axial-vector coupling constant $G_A = \Delta u_0 
- \Delta d_0 = 5/3$ 
and the nucleon spin
$\Sigma = \Delta u_0 + \Delta d _0 = 1$ in the absence of 
the GS boson dressing.  
Inserting these values into eqs.~(\ref{su},\ref{sd},\ref{ss}), 
we obtain
\beqn
\Delta u = 0.86, \hspace{1cm} \Delta d = -0.29, 
\hspace{1cm} \Delta s = -0.006.  
\eeqn
With inclusion of the GS boson dressing the nucleon axial-vector coupling 
constant becomes  
\beq
G_A \equiv \Delta u - \Delta d = 1.15 \; ,
\label{def_ga}
\eeq
in reasonable agreement with the empirical value.    
The total quark fraction of the nucleon spin is then given by
\beq
\Sigma  \equiv \Delta u + \Delta d + \Delta s = 0.56
\label{spin}
\eeq
This value is about twice as large as the empirical $\Sigma = 0.29 \pm 0.06 
$ \cite{Ellis}.  
If we allow ourselves to vary the momentum space cutoff, it would be 
possible to obtain a value around  $\Sigma \sim 0.3$.  
However, the agreement with 
the nucleon axial-vector coupling constant $G_A$ is 
then lost.  We therefore conclude that the nucleon spin 
problem cannot be solved by the GS boson dressing alone, but the 
depolarization caused by the P-wave coupling to 
the GS bosons is nevertheless a significant effect.

The value of $\Sigma $,  
eq.(\ref{spin}), indicates that about 40$\%$ of the nucleon spin is carried by 
the relative orbital angular momentum 
between constituent quarks and their pseudoscalar meson clouds.   
This result is quite consistent with the recent analysis of the nucleon
spin decomposition, according to which about  $\sim 30 \%$ of the 
nucleon spin is carried by
the quark orbital angular momentum \cite{Ji}.    
This apparent agreement is of some interest.

\vspace{2cm}

\ni
{\bf 3.3 Axial anomaly effects}

\ni
Let us now investigate some necessary further steps.  
In the present framework, we deal only with the $SU(3)$ octet of Goldstone 
bosons to build up the 
constituent quark structure; we do not incorporate the contribution 
of the axial $U(1)$ anomaly
to the nucleon spin.  After the EMC, SMC and SLAC measurements, 
many efforts have been made to
connect the missing nucleon spin with the $U(1)_A$ anomaly of QCD.  
Such anomaly contributions to the spin structure of the constituent quarks 
have been estimated \cite{SW} using the three flavour 
Nambu and Jona-Lasinio model 
which dynamically produces spontaneous chiral symmetry breaking and 
incorporates the axial $U(1)$ anomaly in the form of 't~Hooft's effective
interaction between quarks \cite{NJL}.  
Mean field effects from such interactions induce a non-trivial spin structure
of constituent quarks such that the spin fractions change from their 
``bare'' values $\Delta u_0 = 4/3$, $\Delta d_0=-1/3$ and $\Delta s_0 = 0$ 
already before GS boson fluctuations are turned on.   
In a scenario with maximal $\eta$-$\eta'$ mixing and with inclusion of 
relativistic bound state wave functions, Yabu {\etal} find \cite{SW}, 
\beq
\Delta u_0 = 0.91, \hspace{1cm} \Delta d_0 = -0.37, 
\hspace{1cm} \Delta s_0 = -0.12 \;\; .  
\label{input_ano}
\eeq
This result shows the screening of the singlet axial charge induced by the 
quark-antiquark polarization.  
The dressing with pseudoscalar mesons should be viewed as an additional 
effect beyond the mean field level.  
It is therefore meaningful not to start from the standard $SU(6)$ symmetric 
relation, but to adopt the values 
quoted in (\ref{input_ano}).  In this case the strange quarks are 
primordially polarized, and thus the formula obtained previously is slightly
modified.  
We then find a smaller and reasonable value for the nucleon spin as expected.  
\beq
&&\Sigma = 0.20, \neqn \\
&&\Delta u = 0.60, \hspace{1cm} \Delta d = -0.28, 
\hspace{1cm} \Delta s = -0.12\; .
\label{result_ano}
\eeq
However, although the total spin sum 
$\Sigma$ is now consistent with the empirical value 
$0.29  \pm 0.06$\cite{Ellis}, 
the individual spin fractions of $u$- and $d$ quarks disagree with the data, 
and hence 
the resulting nucleon axial-vector coupling constant 
$G_A=0.85$ is about $30 \%$ smaller than the empirical one.

\vspace{2cm}

\ni
{\bf 3.4 Pion contribution to the axial-vector coupling constant}

\ni
At this point, we should emphasize that the pion cloud also 
contributes directly   
to the axial-vector matrix elements illustrated schematically in Fig.3(b), 
in addition to the contributions already discussed 
above (Fig.3(a)) which are nothing but the renormalization and 
depolarization effects due to GS boson emission.  
In fact, within chiral models of the nucleon such as the chiral bag 
model, the contribution from the pion cloud can be explicitly 
calculated\cite{CBM}.  
In the chiral limit, $m_\pi=0$, the pseudoscalar current induced by the 
pion contributes to the axial-vector matrix element due to the zero mass 
pion pole.  
One can show \cite{Jaffe_ax} 
that the magnitude of the pion cloud contribution Fig.3(b) is half  
of the quark one, Fig.3(a).  
The nucleon axial-vector coupling constant eq.~(\ref{def_ga}) is then 
modified  as
\beqn
G_A \, s_\mu &=& \bra P | \bar \psi \gamma_\mu \gamma_5 
{\tau_3} \psi | P \ket\\
&=& (\Delta u - \Delta d) s_\mu \\
&=& [(\Delta u _a + \Delta u_b) - (\Delta d _ a + \Delta d _ b)] s_\mu \\
& \sim & 1.5 (\Delta u_a - \Delta d_a ) \, s_\mu \; ,
\eeqn
where $\Delta q_a$ are the contributions from diagram 3(a), 
essentially the same 
as those calculated in the previous subsection, and $\Delta q_b$ comes from the
pion pole, Fig.3(b).  
Note that this pion effect contributes selectively only 
to the isovector axial-vector matrix elements $G_A^{(3)} \equiv G_A$.  
It does not change the flavor singlet axial current and therefore leaves 
the spin $\Sigma = G_A^{(0)}$ untouched, $\Delta u_b + \Delta d_b=0$.  
Hence this additional effect may well increase $G_A$ to compensate for the 
small value 
obtained in eq.~(\ref{result_ano}), with the total spin being unchanged.  
Indeed, we can obtain the following results including the pion contribution 
of Fig.3(b) in the chiral limit:   
\beqn
&&\Delta u = 0.93,  \hspace{1cm} \Delta d = -0.43, \hspace{1cm}
\Delta s = - 0.12, \\
&&G_A = 1.35, \\
&&\Sigma = 0.21 \; , 
\eeqn
where we have used eq.~(\ref{input_ano}) as input for the bare spin 
distribution 
and introduce the cutoff such as to reproduce the GSR.  
The resulting spin fractions are now reasonably consistent with the empirical
ones, eq.~(\ref{spin_exp}), although 
the axial-vector coupling is somewhat overestimated in the chiral limit, $m_\pi
\ra 0$.

%%%%%%%%%%%%%%%%%%%%%%%%%%%%%%%%%%%%%%%%%%%%%%%%%%%%%%%%%%%%%%%%%%%%%%%%
\begin{figure} 
\begin{center} 
\psfig{file=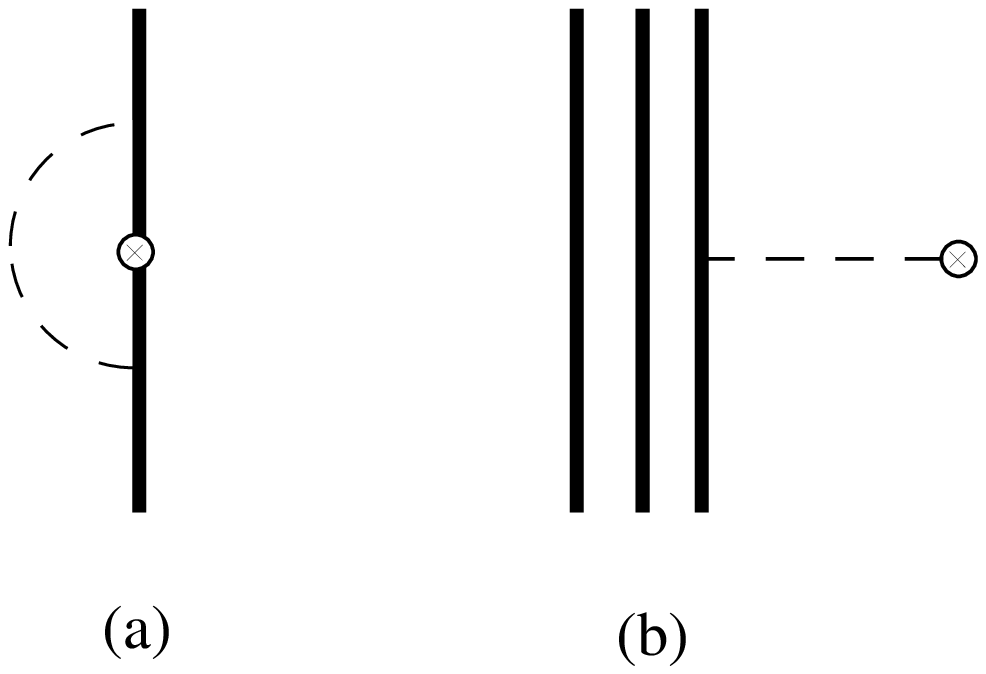,height=1.6in} 
\caption{}
\end{center} 
\vspace{-0.2cm} 
Contributions to the axial-vector matrix element from the pion cloud.  
Notations are the same as those in Fig.1, and the circle denotes the 
insertion of the axial-vector current. The diagram (a) is basically 
the same
as Fig.1(a).  The diagram (b) is the pion cloud effect discussed in the text.  
\end{figure} 
%%%%%%%%%%%%%%%%%%%%%%%%%%%%%%%%%%%%%%%%%%%%%%%%%%%%%%%%%%%%%%%%%%%%%%%%

In the case of the physical pion with 
$m_\pi \sim 140\MeV$, we do not have the soft 
pion pole, but the axial-vector current itself is changed as 
\beq
A^a_\mu = \bar \psi \gamma_\mu \gamma_5 \frac{\tau^a}{2} \psi + 
f_\pi \del_\mu \pi^a \; .
\label{axial_curr_pi}
\eeq
We must rely on some specific model to calculate the second term of  
eq.~(\ref{axial_curr_pi}), which is beyond the purpose of this paper.  
The results are model dependent\cite{axial}, and as a consequence we 
cannot draw strong conclusions about the detailed 
role of such additional pion effects.  
The tendency of the pion cloud to selectively increase the (isovector) 
axial constant $G_A$ is nevertheless obvious.  As a sideremark we note that 
it is also interesting to study the Bjorken-$x$ dependence of the quark
distribution function arising from such pion components, which is expected 
to be centered in the small-$x$ region.

\vspace{2cm}

\ni
{\bf 3.5 Cutoff dependence}

\ni
One might argue that introducing a cutoff as in eq.~(17) implies a high degree
of arbitrariness.  On the other hand, a distinction between ``soft" and ``hard"
scales is needed, the soft physics being represented by the chiral effective 
lagrangian.  Short distance dynamics, not incorporated in the leading order 
effective 
interaction (1), must then reappear in finite size effects parameterized by 
form factors with a cutoff related to the chiral symmetry breaking scale.   
For completeness  
we have investigated other forms of the momentum cutoff function
in order to estimate uncertainties inherent in the cutoff procedure.  
Consider first a sharp cutoff for the transverse 
momentum, $\theta (\Lambda^2 - k_T^2)$.  To reproduce the 
observed value of the Gottfried sum 
the cutoff is the found to be $\Lambda = 0.6 \GeV$. In this case the resulting 
renormalization factor of the constituent quark wave function 
is $Z=0.63$, and the first moments of the splitting functions 
are given by $\bra P_\pi \ket = 0.17$ and $\bra \Delta P_\pi \ket = -0.04$.  
Note that the magnitude of $\bra \Delta P_\pi \ket$ is  smaller 
than the one obtained with a Gaussian cutoff, that is, 
the depolarization effect by coupling to the pion becomes smaller.  
When we adopt the input spin fractions of Yabu {\etal}, eq.~(\ref{input_ano}), 
we find 
$\Delta u = 0.57$, $\Delta d = -0.25$, $\Delta s = -0.11$ and 
$\Sigma = 0.20$, not far from the values, eqs.~(\ref{result_ano}).

We have also used a dipole form for the cutoff,  
$\left( \frac{\Lambda^2 + m_i^2} { \Lambda ^ 2 + M_{j \alpha} ^2 } 
\right)^2 $.  
Here we need  $\Lambda = 2.4 \GeV$ to account for the GSR.  
The resulting renormalization factor is $Z = 0.65$.  
The calculated spin fractions carried by quarks  are 
$\Delta u = 0.58$, $\Delta d = -0.28 $, $\Delta s = -0.12 $ and $\Sigma = 
0.18$.   
As a common feature of all those different cutoff procedures,  we note that the
leading effect on the $\Delta q$'s still comes from wave function
renormalization represented by the $Z$-factor (\ref{Z}).  
We obtain similar values for this $Z$-factor in all cases, once  
the cutoff is fixed to reproduce the GSR.  
From these studies we learn that the cutoff scheme dependence of the results
is within  $10 \%$.

\vspace{2cm}

\ni
{\bf 3.6 Discussion}

\ni
The limitations of the GS boson dressing model to 
explain the nucleon spin structure are obvious.   
First, no antiquark polarization is produced in this approach, because 
the antiquarks are locked into the GS bosons  
which carry no spin.   
However, experimental data indicate that the antiquark polarization is 
not negligible.  
Several lattice calculations also show that the so-called disconnected parts, 
which may involve the effects of the $U(1)_A$ anomaly and 
OZI-violating processes, give 
substantial and negative contributions to the quark 
polarization\cite{Lattice}.  Such contributions are not incorporated in the
diagrams of Fig.1(a) and (b).  
%This is the reason why this model can not reproduce the data for $\Delta u $
%and $\Delta d$.  

Another restriction is the smallness of the strange quark polarization.  
In the original work by Eichten {\etal}, a large strange quark polarization 
was obtained\cite{Eichten}.  
However, once the $SU(3)$ flavor symmetry breaking is taken into account, 
the strange quark polarization from the GS boson couplings turns out to be 
small.  
We recall that the numerator of the spin-dependent splitting function is 
given by $ (m_1 x - m_2)^2 - k_T^2$.  
For $u,d$ quarks with $m_1 = m_2 = m_u$, 
the typical scale of the transverse momentum $k_T$ is larger than the 
constituent quark mass, and thus the splitting function becomes
negative.  On the other hand, in the strange quark production process, 
we have $m_1 = m_u$ and  $m_2 = m_s$.   The large strange quark mass 
cancels the 
negative contribution from the transverse momentum integral.  Therefore, the 
spin-dependent splitting function for the strange quark production becomes  
quite small.

Of course, GS boson dressing is not the only mechanism in question.  We have
pointed out that it should be seen in combination with other screening effects,
involving {\eg} the axial anomaly and OZI-violating processes, which
renormalize the spin structure of constituent quarks as in 
eq.~(\ref{input_ano}).  The sum of all effects has some analogy with the
Arima-Horie renormalization of nucleon $g$-factors in nuclei\cite{Arima}.

It is nevertheless interesting that the GS bosons carry about $35 \%$ 
of the total nucleon  spin.  
Recently, Ji has discussed a gauge invariant decomposition of the 
nucleon spin \cite{Ji}.   
It is suggested that the nucleon spin is carried in parts by the quark spin 
($\sim 20 \%$), the quark (and antiquark) angular momentum ($\sim 30 \%$), 
and the gluons $(\sim 50 \%$) at the low energy scale\cite{Ji}.  
Such a scenario will possibly be tested in future experiments, {\eg} 
by deeply virtual Compton scattering off nucleons.

\vspace{2cm}
%\new
%----------%----------%----------%----------%----------%----------%----------
\ni
{\bf 4 Corrections to the chiral-odd structure function $h_1(x)$}

\ni
We have already discussed the spin-independent quark distribution  
function $f_1(x)$ and 
spin-dependent one $g_1(x)$ in the previous sections.  
We now focus on the chiral-odd transversity  structure
function $h_1(x)$\cite{JaffeJi}.  
The longitudinally polarized structure function $g_1(x)$, which is chiral-even
and can thus be observed in deep inelastic scattering, gives the helicity 
difference of quarks and antiquarks in the longitudinally polarized nucleon.  
On the other hand, 
the chiral-odd $h_1(x)$ structure function corresponds to a target 
helicity-flip amplitude in the standard helicity basis, 
and thus does not have a simple partonic probability   
interpretation.  The $h_1(x)$ structure function provides a correlation
between the left- and right-handed quarks due to its chiral-odd nature.  
%As pointed out in ref.~\cite{JaffeJi}, however,  
%$h_1(x)$ structure function, the transversity spin distribution can 
%be understood as difference of  
%numbers of valence quarks with eigenvalues $+1$ and $-1$ of the transverse 
%Pauli-Lubanski operator $S_T \gamma_5$ in the transversely polarized 
%nucleon.   
Although $h_1(x)$ is not identified with quark spin fraction of
the nucleon, the transversity structure function 
$h_1(x)$ can be understood as the difference of the  
numbers of quarks with eigenvalues $+1$ and $-1$ of the transverse 
Pauli-Lubanski operator $S_T \gamma_5$ in the transversely polarized 
nucleon\cite{JaffeJi}.   
Unlike the axial charge of the $g_1(x)$ distributions, the gluon operators 
which contribute to $h_1(x)$ 
do not mix quark operators under renormalization because of the chirality.  
The analysis of $h_1(x)$ therefore gives complementary information on the
nucleon spin structure.

Naive non-relativistic quark models predict $f_1(x)=g_1(x)=h_1(x)$.  
Relativistic quark models such as the MIT bag model 
give a relation $ | h_1(x)| \ge | g_1(x)|$ due 
to the role of the lower components of the quark 
Dirac wave function\cite{JaffeJi}.  
In such models, the following simple expressions are obtained for the first 
moments of $g_1(x)$ and $h_1(x)$, the {\em axial charge} $\Delta q$ and 
{\em tensor charge} $\delta q$: 
\beqn
| \Delta q | &\sim& \int d^3 r [F^2 - \frac{1}{3}G^2] \; ,\\
| \delta q | &\sim& \int d^3 r [F^2 + \frac{1}{3} G^2] \; ,
\eeqn
where $F$ and $G$ are the upper and lower components of the quark 
Dirac spinor.  
Recent lattice simulations indicate that 
the tensor charge is larger than
the axial charge for the $u$-quark,  whereas the magnitude of the 
tensor charge becomes smaller than the axial one for the 
$d$-quark \cite{h1lattice}, results 
which disagree with simple bag model expectations.   
Here, we study how the pion dressing modifies the $h_1(x)$ spin structure
function.

Since the $h_1(x)$ structure function is chiral-odd, it cannot be 
measured by a chiral-even probe such as the electro-magnetic interaction.  
Following the work of Ioffe and Khodjiamirian\cite{Ioffe}, we introduce the 
chiral-odd forward scattering amplitude  
\beq
T_\mu (q,p,s)= \frac{i}{2} \int { d^4 \xi e^{iq\cdot \xi } \left\langle 
{p,S|T \{ J_{\mu 5}(\xi ) \, J(0) + J(\xi ) \, J_{\mu 5}(0) \} |p,S}
 \right\rangle } \; , 
\eeq
where  $J_{\mu 5} = \bar \psi \gamma_\mu \gamma^5 \psi$ is the 
axial-vector current and $J=\bar \psi \psi$ the scalar current.  
This amplitude is related with the chiral-odd 
structure function $h_1(x)$ through the optical theorem.  One can write 
\beq
T_\mu (q,p,s)=\left( {s_\mu -{{p\cdot q} \over {q^2}}q_\mu } 
\right) \tilde  h_1(x)+ \cdots 
\eeq
and obtains the transversity spin distribution function as 
\beq
h_1(x) = - \frac{1}{\pi} \mbox{Im} \tilde h_1(x)
\eeq
We use the same techniques as in Section 2 to evaluate the GS boson correction
to $h_1(x)$.  The splitting function for the transversity difference 
is given by,
\beq
\delta P(y)_{j \, \alpha / i}={1 \over {8\pi ^2}}\left( {{{g_A \bar m} 
\over {f_\pi }}} \right)^2\int_{}^{} {dk_T^2}
{{(m_j - m_i y)^2 } \over {y^2 (1-y) 
\left[ {m_i^2-M^2_{j\, \alpha}} \right]^2}}
\label{spil-h}
\eeq
This result tells us that a relation 
\beq
P(x) + \Delta P(x) = 2 \delta P(x)
\eeq
holds among the GS boson corrections, which implies a saturation of  
Soffer's inequality \cite{Soffer} in the chiral quark model.  
The calculated splitting function $ \delta P_{j \, \alpha /i}$ 
is shown in Fig.2 together with $ P_{j \, \alpha /i}$ 
and  $\Delta P_{j \, \alpha /i}$.  
Note that $\delta P  (x)$ is small and positive in the whole $x$ region, 
in contrast to the negative longitudinal
splitting function $\Delta P (x)$.  
The first moment of the transversity splitting function is 
\beqn
\left\langle {\delta P_\pi } \right\rangle = 0.05 \; ,
\eeqn
to be compared with $\left\langle { P_\pi } \right\rangle = 0.16$ 
and  $\left\langle { \Delta P_\pi } \right\rangle = -0.06$.  
The sign difference between $\delta P$ and $\Delta P$ 
causes a non-negligible modification for the $d$-quark distribution 
functions.  Since the numerator of the transversity splitting function is 
independent of
the transverse momentum, $\bra \delta P \ket $ is insensitive to the 
choice of the momentum cutoff.

In order to estimate the GS boson corrections explicitly, 
we need the bare quark distributions as inputs.  
Here and for convenience, we use a covariant quark-diquark model, 
which has already been studied in the 
literature\cite{scalar,Artru,Suzuki}\footnote{In ref.~\cite{Kulagin} 
both scalar and axial-vector quark-diquark nucleon vertices are incorporated 
to calculate the unpolarized structure function.  Here we take 
only the scalar vertex for simplicity, although we could 
use the same procedure to calculate the spin-dependent distributions.  
Hence model parameters used here differ 
from those in ref.~\cite{Kulagin}.  Our main concern here is to
clarify effects of the Goldstone boson dressing 
on quark distributions.  These effects are 
independent of minor differences in the input quark distributions.}  
to obtain the input distributions  for $f_1(x)$, $g_1(x)$ and $h_1(x)$.

Assuming $SU(4)$ spin-flavor symmetry, quark distribution functions 
of bare constituent quarks can be written as
\beqn
u_0 (x) &=& 2 S_f (x), \hspace{1cm} d_0(x) = S_f(x) ,  \\
\Delta u_0 (x) &=& \frac{4}{3} S_g(x), 
\hspace{1cm} \Delta d_0 (x) = -\frac{1}{3} S_g(x) ,  \\
\delta u_0 (x) &=& \frac{4}{3} S_h(x), \hspace{1cm}
\delta d_0 (x) = - \frac{1}{3} S_h(x) .  
\eeqn
Expressions for $S_f (x), S_g(x), S_h (x) $ are given explicitly in 
refs.~\cite{scalar,Artru,Suzuki}.   
The relation $S_h (x) > S_g (x)$  
is satisfied as in the bag model calculation\cite{JaffeJi}.  
We do not have to elaborate the input distributions here, 
since the essential features of the following results are
independent of the shape of these functions.

As a consequence of the GS boson dressing, the spin-dependent   
distribution functions for $u$-and $d$-quarks are modified in the
following way: 
\beq
\Delta u (x) &=& Z \Delta u_0(x) +\frac{1}{2} \Delta P_\pi \otimes 
\Delta u_0 + 
\Delta P_\pi \otimes \Delta d_0 + \frac{1}{6} \Delta P_\eta  
\otimes \Delta u_0  \neqn \\
&=&  \frac {4} {3} Z S_g + \frac{1}{3} \Delta P_\pi \otimes S_g + 
\frac{2}{9} \Delta P_\eta \otimes S_g  \; ,
\label{g1_u} 
\eeq
\beq
\delta u (x) &=&  \frac {4} {3} Z S_h + \frac{1}{3} \delta P_\pi \otimes S_h + 
\frac{2}{9} \delta P_\eta \otimes S_h \; ,
\label{h1_u}
\eeq
\beq
\Delta d (x) &=& Z \Delta d_0(x)+ \frac{1}{2} \Delta P_\pi \otimes 
\Delta d_0 + 
\Delta P_\pi \otimes \Delta u_0 + \frac{1}{6} \Delta P_\eta  \otimes 
\Delta d_0 \; , \neqn \\
&=&  -\frac {1} {3} Z S_g + \frac{7}{6} \Delta P_\pi \otimes S_g 
- \frac{1}{18} \Delta P_\eta \otimes S_g \; ,
\label{g1_d}
\eeq
\beq
\delta d (x) &=& -\frac {1} {3} Z S_h + \frac{7}{6} \delta P_\pi \otimes S_h  
- \frac{1}{18} \delta P_\eta \otimes S_h 
\label{h1_d}
\eeq
%
%
%It is easily seen from eqs.(\ref{g1_u},\ref{g1_d},\ref{h1_u},\ref{h1_d}) 
%that the corrections are considerable to the 
%$u$-quark distribution due to simple spin-flavor coefficents, 
%while they are small for the $u$-quark.  

\ni
We show in Figs.4  the spin-dependent $u$-quark distributions in the proton,  
$\Delta u(x)$ and $\delta u(x)$,  together with the unpolarized distribution 
$u(x)$.   Results with inclusion of GS boson fluctuations are given  
by the solid curves, the  ones without the GS bosons by the dashed lines.  
The renormalization of the constituent quark wave function reduces 
these distribution functions from their original distributions, 
and the small-$x$ region is enhanced by the GS bosons.  
However, the relative magnitudes of $u(x)$, $\Delta u(x)$ and $\delta u(x)$ 
are not modified very much.   
The first moments of $\Delta u(x)$ and $\delta u(x)$ 
are shown in Table 1 with corresponding moments of the 
original distribution functions.

%%%%%%%%%%%%%%%%%%%%%%%%%%%%%%%%%%%%%%%%%%%%%%%%%%%%%%%%%%%%%%%%%%%%%%%%
\begin{figure} 
\begin{center} 
\psfig{file=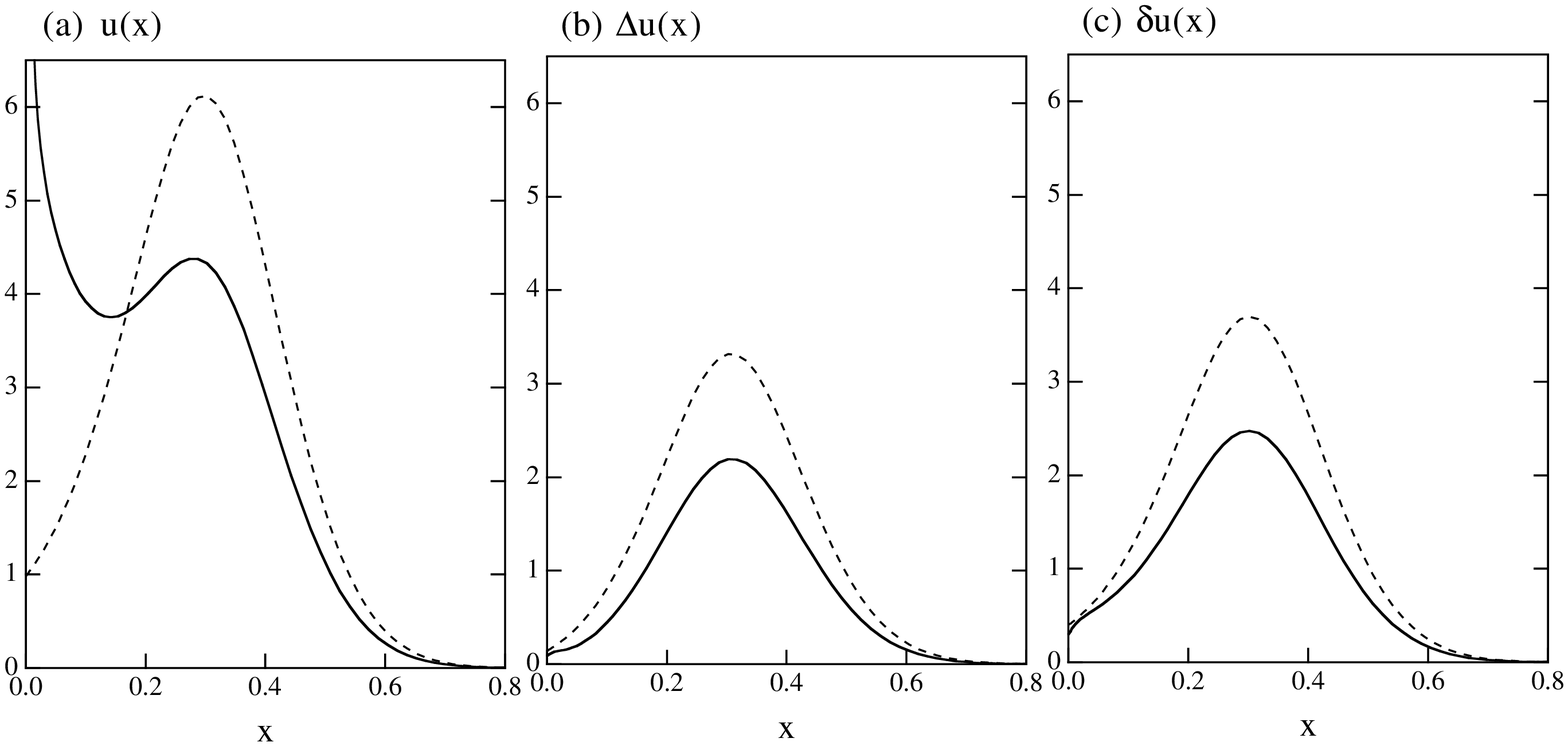,height=2.65in} 
\caption{}   
\end{center} 
\vspace{-0.2cm} 
The $u$-quark distribution functions:   
(a) $u (x)$, (b) $\Delta u(x)$ and (c) $\delta u(x)$, respectively.  
In each figure, the result with dressed constituent quarks is shown by the
solid curve, the one without the dressing by the dashed curve.  
Here we use the following parameters to obtain the input bare distributions
(dashed): quark mass 0.36GeV, diquark mass 0.7GeV 
and a cutoff 0.8GeV in the quark-scalar diquark model\cite{Suzuki}.  
Contributions from the pion tail Fig.3(b) to the axial charge are not
included.  
\end{figure} 
%%%%%%%%%%%%%%%%%%%%%%%%%%%%%%%%%%%%%%%%%%%%%%%%%%%%%%%%%%%%%%%%%%%%%%%%

Effects of the GS boson dressing are more apparent in the 
$d$-quark case shown in Figs.5.  
The original distributions satisfy $| \delta d_0 (x)|  \ge 
| \Delta d_0 (x) | $.  
Once the dressing corrections are turned on, $| \delta d_0 (x) | $  becomes
considerably smaller.  
In the $\Delta d(x)$ distribution function, 
the corrections from the
renormalization and $\bra \Delta P \ket$ cancel each other, and  
the resulting $\Delta d(x)$ is not much modified.  On the other hand, 
both contributions are positive for $\delta d (x)$, 
and hence the transversity of the $d$-quarks is reduced drastically.

%%%%%%%%%%%%%%%%%%%%%%%%%%%%%%%%%%%%%%%%%%%%%%%%%%%%%%%%%%%%%%%%%%%%%%%%
\begin{figure} 
\begin{center} 
\psfig{file=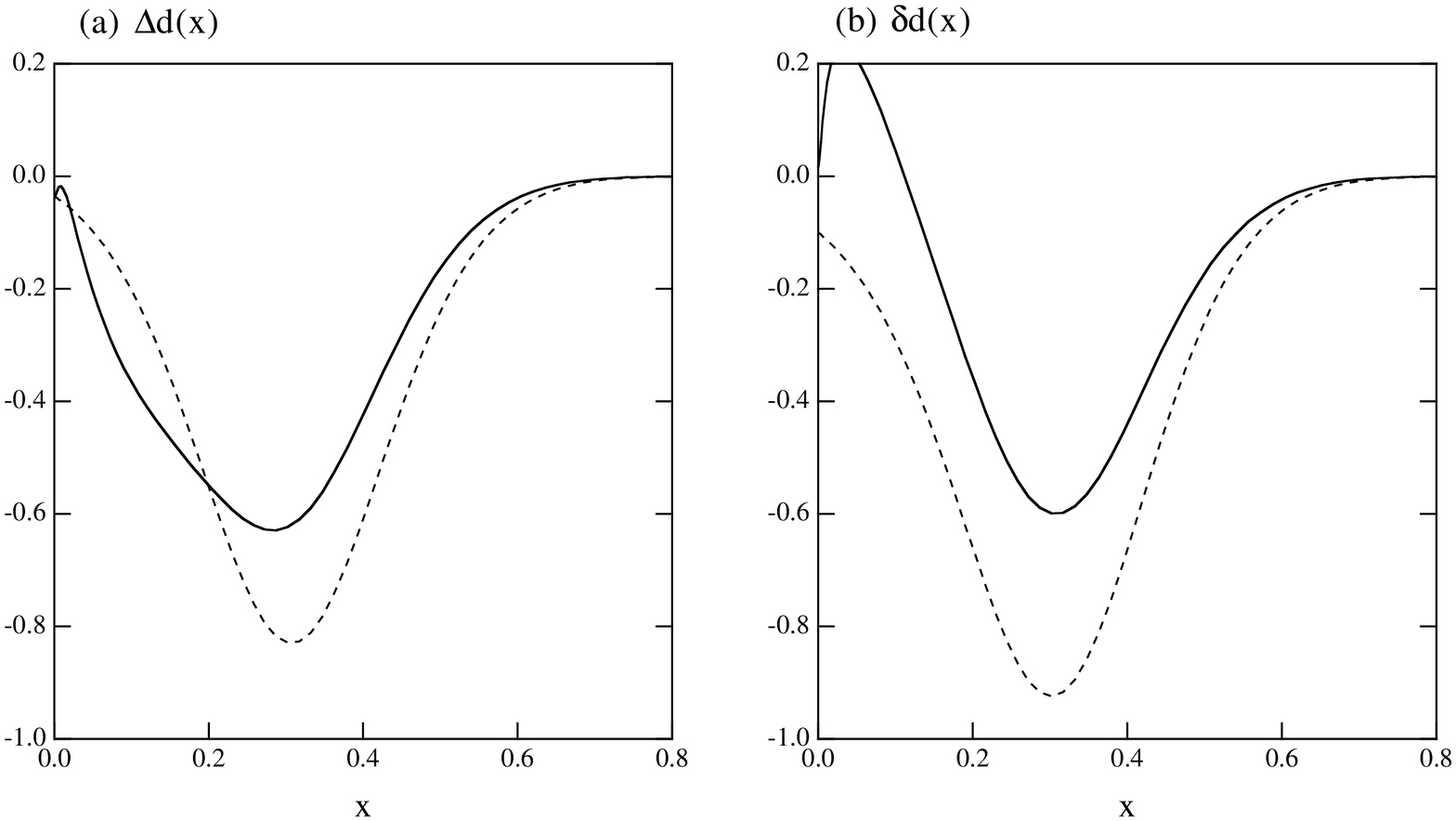,height=3in} 
\caption{}   
\end{center} 
\vspace{-0.2cm} 
The $d$-quark distribution functions: 
(a) $\Delta d (x)$ and (b) $\delta d (x)$, respectively.  
Notations are the same as those of Fig.4.
\end{figure} 
%%%%%%%%%%%%%%%%%%%%%%%%%%%%%%%%%%%%%%%%%%%%%%%%%%%%%%%%%%%%%%%%%%%%%%%%

The first moments are tabulated in Table 1.  
The original quark-diquark model without the CQ internal structure predicts 
the universal inequality  $ | \Delta q | < | \delta q |$.  
The GS boson dressing then changes the moments of the $d$-quark, and 
$ | \Delta d | > | \delta d |$ is obtained.   
We find that the $d$-quark tensor charge is reduced by about 
$40 \sim 50 \%$.  
%If we used more elaborate model of the nucleon, the resulting axial charge 
%could become closer to the experiments.  
This result is essentially parameter independent and applies for 
any model calculation.  
We note that the smallness of the $d$-quark tensor charge is also obtained 
within the 
QCD sum rule approach.  In recent work He and Ji find $\delta u = 
1.33 \pm 0.55$ and $\delta d = 0.04 \pm 0.02$ \cite{QSR}.

\vspace{0.5cm}

{\center

{\bf Table 1}

\vspace{0.5cm}

\begin{tabular}{l|c|c|c|c}
\hline
        & $\Delta u $   & $\Delta d $   & $\delta u $  & $\delta d$\\ \hline
Bare    & 1.01          &$-0.25$        & 1.17         & $-0.29$   \\ 
With CQ & 0.65          &$-0.22$        & 0.80         & $-0.15$   \\ 
Exp.    & $0.82\pm 0.02$ & $-0.43\pm0.02 $&   $- $       &   $-$     \\
Lattice &  0.76         & $-0.35$     & 0.84        
 & $-0.23$   \\ \hline
\end{tabular}

}

\vspace{0.5cm}

\baselineskip = 0.45cm
\ni
The first moments of the helicity and transversity distribution 
functions.  Results with bare quarks and dressed quarks are shown in the 
second and third columns, respectively.  
Experimental data\cite{Ellis} and the lattice simulations\cite{h1lattice} 
are also shown in the fourth and fifth columns.  
Here we use the following parameters: quark mass 0.36GeV, diquark mass 0.7GeV 
and a cutoff 0.8GeV in the quark-scalar diquark model\cite{Suzuki}.  
Contributions from the pion tail Fig.3(b) to the axial charge are not
included.  

\baselineskip = 0.765cm

\vspace{0.5cm}

It is worthwhile mentioning here that the disconnected
contribution in the lattice QCD simulation 
to $h_1(x)$ is negligibly small.   This situation is quite different
from the helicity distribution $g_1(x)$, for which the disconnected parts 
are sizeable.

We comment on recent developments in the perturbative evolution of the 
$h_1(x)$ structure function\cite{h1_spil}.  
The Altarelli-Parisi splitting kernel for $h_1(x)$ is different from the 
standard one for the spin structure function.  
Barone {\etal}\cite{Barone} found that, even if $h_1(x)$ and $g_1(x)$ are 
similar at a low starting scale $\sim 0.5 \GeV$ as suggested by 
simple quark models, 
$h_1(x)$ becomes much smaller than $g_1(x)$ for  small $x < 0.1$ 
at the experimentally accessible scale due to the difference in the $Q^2$
evolution.  
On the other hand, the GS boson cloud effects studied in the present paper 
show up 
at intermediate values of the Bjorken variable, $0.1 < x < 0.5$.  
In particular 
the $d$-quark contribution to $h_1(x)$ is much reduced 
around $x =0.2 \sim 0.3 $ from the 
simple quark model estimate.  The GS boson effect also reduces the $d$-quark
tensor charge very much as already pointed out.   
We conclude that the GS boson
cloud effects, if existent, should be seen experimentally in the 
intermediate $x$ region.

In this paper we omit the Regge-type contributions at small-$x$,
although significant effects for the unpolarized structure function were 
found in ref.~\cite{Kulagin}.  It is clearly an interesting point, 
still under investigation,  
to combine the GS boson cloud effects presented here and Regge phenomenology 
in order to reach a unified picture of non-perturbative features at all $x$.  
Small-$x$ behaviour was studied for $g_1(x)$\cite{smallg1} and 
$h_1(x)$\cite{smallh1} within the leading logarithmic approximation of 
perturbative QCD, and interesting differences were found.

\vspace{2cm}

%\new
%%%%%%%%%%%%%%%%%%%%%%%%%%%%%%%%%%%%%%%%%%%%%%%%%%%%%%%%%%%%%%%%%%%%%%%%
\ni
{\bf 5 Sea quark distributions in the pion and the nucleon}

\ni
In the framework of the chiral quark model, constituent quarks and Goldstone 
bosons are treated as quasi-particles.   
They appear structure-less at the tree-level.  
As shown in previous sections, however, the dressing of quarks plays 
an important role in understanding the nucleon structure.  
Here we try to describe the pion in the similar way, and see how the pion 
structure
function changes when including higher order corrections derived from the
effective Lagrangian of the chiral quark model.  
We specially concentrate on the sea quarks 
in the nucleon and the pion and show that the sea quark momentum 
fraction in the pion is expected to be significantly larger than that of 
the nucleon.

%%%%%%%%%%%%%%%%%%%%%%%%%%%%%%%%%%%%%%%%%%%%%%%%%%%%%%%%%%%%%%%%%%%%%%%%
\begin{figure} 
\begin{center} 
\psfig{file=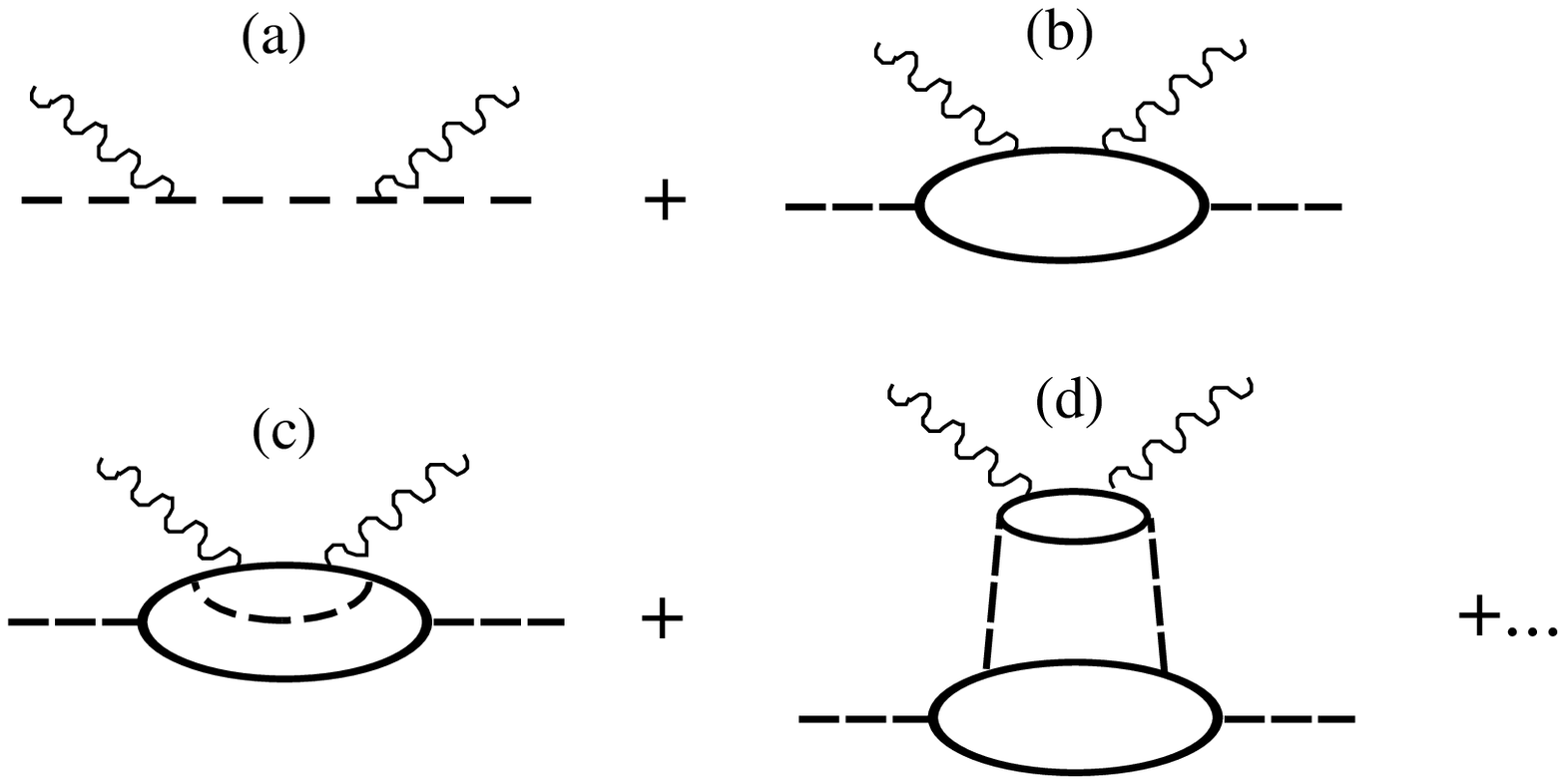,height=2.7in} 
\caption{}   
\end{center} 
\vspace{-0.2cm} 
Typical diagrams contributing to the substructure of the pion in the chiral 
quark model 
\end{figure} 
%%%%%%%%%%%%%%%%%%%%%%%%%%%%%%%%%%%%%%%%%%%%%%%%%%%%%%%%%%%%%%%%%%%%%%%%

Typical diagrams contributing to the pion-virtual photon forward scattering
amplitude are displayed in Fig.6.    
The first diagram, Fig.6(a), represents the elementary 
pion process ({\ie} elastic Compton scattering on the pion) and does not 
contribute to the leading twist distribution.  
The second graph, Fig.6(b), can be calculated within the CQM as a first 
non-trivial contribution.   
We denote this distribution function by $V_{q / \pi} (x) $ and associate it
with the valence quark and antiquark content of the pion.  
The third and forth diagrams, Fig.6(c)-(d), correspond to the case in which 
the quark or the 
antiquark in the second diagram is dressed by the Goldstone bosons.  
These contributions can be evaluated by using the procedure developed in 
the Section 2.  
Such processes determine the sea quark distribution of the pion 
in the chiral quark model.

We have already given expressions for the quark distribution of the nucleon 
in eqs. (\ref{u_N},\ref{d_N},\ref{baru_N},\ref{bard_N}).  
The sea quark distribution in the nucleon is
\beq
S_{N} (x) = 2 \bar u(x) + 2\bar d(x) + s(x) + \bar s(x) \; .
\eeq
Such sea quark distribution functions arise from the diagram, Fig.1(b), 
in which the constituent quarks act as spectators.  
The contribution from this diagram is soft and concentrated in the small 
$x$ region as a result of the double convolution.

On the other hand, for the $\pi^+$ case in which $u_0$ and $\bar d_0$ 
are bare quark distributions, 
the non-valence components that arise from the  dressing are given by
\beq
\bar u(x) &=&  P_\pi \otimes \bar d_0 + 
 {1 \over 4}V_{u / \pi }\otimes P_\pi \otimes (u_0+ \bar d_0)+ 
\frac{1}{36} V_{u  / \eta }\otimes P_\eta \otimes ( u_0+ \bar d_0)  \; ,
\label{ubar_pi}
\eeq
\beq
d (x) &=&  P_\pi \otimes u _0 + 
 {1 \over 4}V_{d / \pi }\otimes P_\pi \otimes (u_0+ \bar d_0)+ 
\frac{1}{36} V_{d  / \eta }\otimes P_\eta \otimes ( u_0+ \bar d_0)  \; ,
\label{d_pi}
\eeq
\beq
s(x) &=& P_K \otimes u _0 + V_{s / K }\otimes P_K \otimes \bar d _0 + 
\frac{4}{9} V_{s / \eta}\otimes P_\eta \otimes (u_0 + \bar d_0)   \; , \neqn \\
\bar s(x) &=& P_K \otimes \bar d _0 + V_{s / K }\otimes P_K \otimes u _0 +
 \frac{4}{9} V_{s / \eta}\otimes P_\eta \otimes (u_0 + \bar d_0)  \; .  
\label{s_pi}
\eeq
For the bare quark distribution we have $u_0  (x) = \bar d_0 (x) = 
V_{q / \pi}(x) $ by definition.  
Thus, one can write the sea quark distribution in the positively charged 
pion as 
\beq
S_\pi (x) = 2 \bar u(x) + 2 d(x) + s(x) + \bar s(x) \; .  
\eeq
Note that both diagrams 1(a) and 1(b) contribute to the sea quark 
distributions in the meson case because of its $q \bar q$ structure.    
The diagram, Fig.1(a), gives harder distribution functions.  
This leads to an interesting difference between the nucleon and the pion.

Let us consider the momentum fraction carried by the sea quarks.  
From eqs.(\ref{baru_N}, \ref{bard_N}) and 
(\ref{ubar_pi},\ref{d_pi},\ref{s_pi}), 
the second moments of the sea quark distributions in the nucleon and
the pion are expressed as
\beq
\bra x S_N \ket  = 2  \bra  x V_{q / \pi} \ket_\pi  \bra x P_{\pi \, j / i} 
\ket 
\bra x (u_0 + d_0) \ket_N  + \cdots \; ,
\label{seamom_N}
\eeq
\beq
\bra x S_{\pi} \ket = 2 \bra x P_{j \pi / i} 
\ket \bra 2 x V_{q / \pi} \ket_\pi + \cdots \; ,
\label{seamom_pi}
\eeq
where we only write down the dominant contributions.  
Here $\bra 2 x V_{q / \pi} \ket_\pi$ and $\bra x (u_0 +  d_0)\ket_N $ 
are the second moments of the bare valence distributions in the 
pion and nucleon, respectively.

One can establish a simple relation 
between the GS-boson-induced sea quark distributions in the nucleon and 
the pion.  
Neglecting the non-leading terms in (\ref{seamom_N}) and (\ref{seamom_pi}),
which are about $20 \%$  of the first terms, 
and taking a ratio, we obtain 
\beq
\frac{\bra x S_{\pi} \ket  }{\bra x S_N \ket} &\sim& 
\frac{ 2 \bra x P_{j \pi / i} \ket} 
{\bra x P_{\pi \, j / i} \ket \bra x (u_0 + d_0) \ket_N  } \neqn \\
& \simeq & \frac{1.46} { \bra x (u_0 + d_0) \ket_N } \; \; ,
\label{rel_sea}
\eeq
where we use the CQM parameters determined in Section 3.   
This approximate relation indicates that the sea quark momentum fraction
in the pion is substantially larger than that in the nucleon,  
because $\bra x (u_0 + d_0) \ket_N$ must be
less than 1.

Here we simply take $ \bra x (u_0 + d_0) \ket_N \sim 0.6$ from the
parametrization of the nucleon structure function \cite{Gluck} at the scale 
$Q^2 = 1 \GeV^2$, at which this relation is assumed to be  
meaningful.  Note that $S_{\pi} $ and $ S_N $ are the sea quark 
distributions generated by the GS boson dressing alone, and cannot be simply
identified with the sea quarks of the standard parametrization which may also  
include sea partons from the perturbative gluon radiation.  
In ref.~\cite{Gluck} the sea quarks in the nucleon carry the momentum fraction 
$ \sim 0.15$ at $1 \GeV^2$.  
As an example, suppose that half of this value  
is generated by the GS boson dressing, {\ie} $\bra x S_N  \ket \sim 0.7$.   
Then we get from eq.~(\ref{rel_sea}),
\beqn
\bra x S_{\pi} \ket \sim 1.7  \; \; , 
\eeqn
a large contribution.  
The total sea quark momentum fraction in the pion 
combined with the sea quarks generated by the
gluon radiation may be more than 0.2.  
Such a large sea quark distribution cannot be generated perturbatively at 
$Q^2 \sim 1 \GeV^2$.  The systematic analysis of the sea quarks in the 
pion might be a key 
to understand the non-perturbative structure of the constituent quarks.

Even if sea partons exist in the nucleon and/or pion 
before the GS boson dressing, the relation (45) remains qualitatively 
unchanged.  
For instance, in such a case, eq.~(\ref{seamom_N}) is rewritten as
\beqn
\bra x S_N \ket  = 2  \bra  x V_{q / \pi} \ket_\pi  \bra x P_{\pi \, j / i} 
\ket \bra x (u_0 + d_0) \ket_N  + 
3 \bra x P_{j \, \pi / i} \ket \bra x (\bar u_0 + \bar d_0) \ket_N + 
\cdots \;  . 
\eeqn
The contribution from the second term is estimated to be $20 \% $ at most 
of the
first term according to the parametrization of the antiquark 
distribution\cite{Gluck}, although the second term arises from the process 
of Fig.1(a).

At present, the sea quark distribution in the pion is not determined 
experimentally.  In the analysis of ref.~\cite{SMRS}, 
both small ($10 \%$) and large ($20 \%$)  sea quark 
distributions in the pion
lead to an equally good fit to the Drell-Yan data.  
Future experiments should clarify this uncertainty.  
A precise determination of the pion sea quark distribution may be possible 
by using the pion-deuteron Drell-Yan experiment\cite{piDY}.

\vspace{2cm}

%-----------%-----------%-----------%-----------%-----------%-----------
%\new
\ni
{\bf 6 Summary and discussions}

\ni
We have studied the Goldstone boson dressing of constituent quarks 
in the chiral $SU(3)$ quark model 
and its implications for nucleon and pion structure functions.  
The GS boson contributions, in which the pion is dominant, produce  
substantial renormalization effects of the constituent quark properties.  
Such a renormalization yields the asymmetry of the $u$ and $d$ quark
distribution in the nucleon.  The resulting value for the Gottfried sum
is consistent with the observed data.   
Similar results are obtained in hadronic model calculations with the meson
clouds\cite{Review_IMF}.

The CQ spin structure is also directly modified by the GS boson fluctuations.  
They change the CQ spin structure by emitting the GS boson in a 
$P$-wave relative to the quark.   
But here the
situation is more subtle.  The spin-dependent splitting function $\Delta P(x)$
is sensitive to the value of the momentum cutoff used in the chiral quark 
model.  
We have fixed this cutoff to reproduce the empirical Gottfried sum, 
and found that the spin fractions of the dressed CQ 
are reduced by about $10$-$30\%$ from the ones of the undressed CQ.  
This depolarization effect is a consequence of spin being converted to P-wave
orbital angular momentum of the meson cloud surrounding the quarks.  
It is difficult, however, to decrease the spin fraction further, when we keep 
simultaneously 
the nucleon axial-vector coupling constant $G_A$ at its measured value.  
Hence we conclude that the nucleon spin problem is not solved by the GS boson
dressing alone.  On the other hand, the 
calculated depolarization in the present study (30-40$\%$) is quite 
consistent with the recent decomposition of the nucleon spin\cite{Ji} 
when identified with the contribution from the quark orbital angular 
momentum.  
When combined with other contributions, namely the axial $U(1)$ anomaly and 
the meson cloud effects illustrated in Fig.3(b), to be 
identified with the disconnected contributions on the lattice QCD simulations, 
the present results go altogether 
into the right direction of providing a reasonable
description of the nucleon spin structure.

We briefly comment on additional corrections from 
the multi-pion cloud of the constituent quark.  
Consider the following two-pion Fock space component of the constituent 
$u$-quark:    
\beq
\left| U_{2 \pi} \right\rangle  &=& 
a_1 \left| {u \pi ^+ \pi^- } 
\right\rangle +  a_2\left| {u \pi ^0 \pi^0 } \right\rangle +
a_3 \left| {d \pi^+ \pi^0} \right\rangle \; \; .  
\eeq
Such contributions are described by the diagram, Fig.7.   
Here we incorporate the ladder diagrams and neglect the crossed pion loop 
diagrams, which give much softer contributions to the structure 
function \cite{Drell}.

%%%%%%%%%%%%%%%%%%%%%%%%%%%%%%%%%%%%%%%%%%%%%%%%%%%%%%%%%%%%%%%%%%%%%%%%
\begin{figure} 
\begin{center} 
\psfig{file=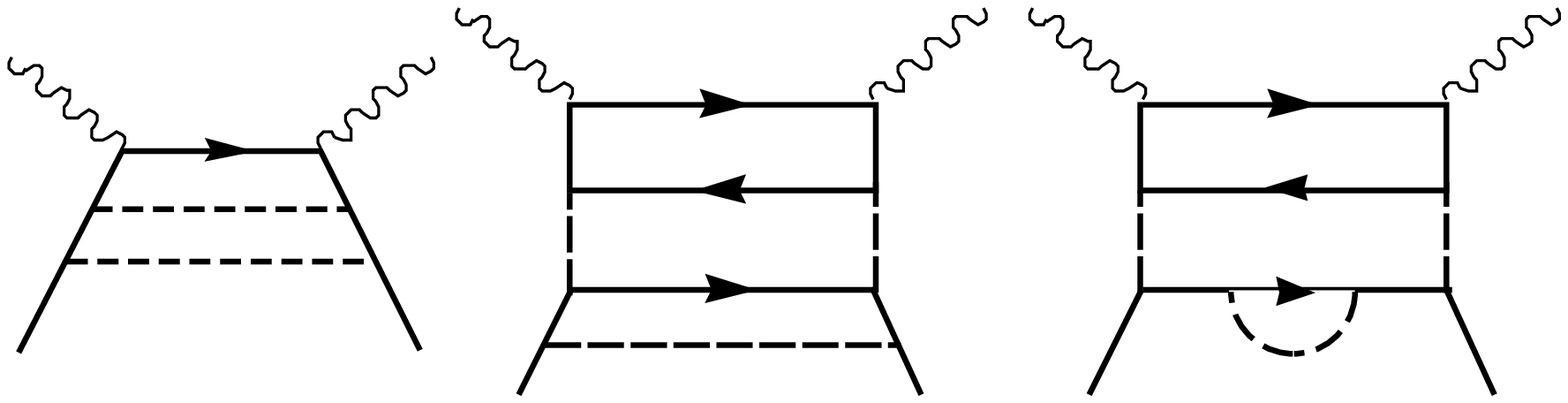,height=1.5in} 
\caption{}   
\end{center} 
\vspace{-0.2cm} 
Typical diagrams contributing to the constituent quark structure function 
with the two GS bosons.  Notations are the same 
as those of Fig.1.
\end{figure} 
%%%%%%%%%%%%%%%%%%%%%%%%%%%%%%%%%%%%%%%%%%%%%%%%%%%%%%%%%%%%%%%%%%%%%%%%

Due to factorization in the IMF, corrections from such diagrams are 
expressed by the already known splitting
functions in most cases.   
We find the following two-pion contribution to the renormalization constant: 
\beq
Z_{2 \pi} =  0.03 \; ,
\eeq
which is about 10$\%$ of the total single GS boson contribution.  
We conclude that 
the inclusion of multi-pion Fock states is not expected to change our 
results significantly.   
The perturbative treatment of the GS boson cloud should therefore 
be justified.

We have also evaluated the GS boson correction to the chiral-odd 
$h_1(x)$ structure function.   In comparison with the result for
$g_1(x)$, where 
the CQ-GS boson splitting function $\Delta P$ is negative, we find a positive
contribution to the $h_1(x)$ structure function.  
With inclusion of the renormalization effect, the transversity spin structure
function for the $d$-quark is shown to be 
reduced significantly.  
We have found the relation $| \delta d | < | \Delta d |$, which seems to 
agree with recent lattice QCD results\cite{h1lattice}.

One may wonder how the QCD evolution of the tensor charge influences 
this discussion.  
In fact, the tensor charge decreases by the QCD 
evolution\cite{Artru,h1_spil}.  
However, even if one evolves from the very low scale 
$\mu^2 = 0.2 \GeV^2$ to a few $\GeV^2$, the reduction of the tensor charge is
at most $10\%$.    We may therefore expect that the evolution does not 
change the relative magnitude of $\Delta q$ and $\delta q$ very much.  
We need some other dynamical mechanism to get $\Delta u < \delta u$ and 
$| \Delta d | > | \delta d|$, such as the GS boson dressing of the constituent 
quark, if the lattice result is confirmed.

Another interesting result has been obtained for the pion structure function.  
Sea quarks in the pion 
appear by direct emission of the GS bosons, while the nucleon case
requires the second order process illustrated in Fig.1(b).  
This picture naturally gives a substantial enhancement of the sea quark 
distribution function in the pion.  
Such an enhancement can be checked in a future Drell-Yan 
experiment\cite{piDY}.    
Our model calculations are done at the low energy scale $\sim 1 \GeV^2$, 
where the chiral quark model is supposed to make sense.    
To make a prediction for forthcoming experiments, we must carry out the 
$Q^2$ evolution from the low energy scale.  
However, the singlet evolution equation needs the gluon momentum fraction 
in the pion at 
the model scale, which we cannot estimate reliably at the moment.

We have studied aspects of constituent quark structure in deep inelastic 
processes.  
Non-perturbative features of the constituent quark, {\ie}~Goldstone boson 
fluctuations around the CQ, lead to screening effects and to 
characteristic changes of the 
CQ spin structure.  Experimental tests of the nucleon transversity spin 
distribution function and the pion sea quark distribution will 
provide new insights into the role of the constituent quarks  
as quasi-particles.

In closing we mention that, recently, considerable interest has been focused 
on the possible experimental observation of such a ``soft" pion cloud 
surrounding the nucleon.   Semi-inclusive deep inelastic scattering processes 
with pions, nucleons or $\Delta $-isobar in the final state are 
studied as promising options by several authors\cite{Semi}.

\vspace{0.5cm}

\ni
{\bf Acknowledgments}

K.S.~would like to thank A.~Hosaka for valuable conversations about the 
contribution of the pion cloud to the axial-vector coupling.    
This work is supported in part by the Alexander von Humboldt foundation 
and by BMBF.

%%%%%%%%%%%%%%%%%%%%%%%%%%%%%%%%%%%%%%%%%%%%%%%%%%%%%%%%%%%
%%%%%%%%%%%%%%%%%%%%% REFERENCES %%%%%%%%%%%%%%%%%%%%%%%%%%
%%%%%%%%%%%%%%%%%%%%%%%%%%%%%%%%%%%%%%%%%%%%%%%%%%%%%%%%%%%

%
%
\end{document}